\documentclass[fleqn,usenatbib]{mnras}
\usepackage{newtxtext,newtxmath}

\usepackage[T1]{fontenc}
\usepackage{ae,aecompl}

\usepackage{graphicx}	
\usepackage{amsmath}	
\usepackage{amssymb}	

\newcommand{\Msun}{\ensuremath{\text{M}_\odot}}
\newcommand{\Msunyr}{\ensuremath{{\text{M}_\odot/\text{yr}}}}

\newcommand{\SFR}{\ensuremath{SFR}}
\newcommand{\sSFR}{\ensuremath{sSFR}}
\newcommand{\Mstar}{\ensuremath{M_*}}
\newcommand{\lsSFR}{\ensuremath{{\log(\frac{\sSFR}{\text{yr}^{-1}})}}}
\newcommand{\lSFR}{\ensuremath{{\log(\frac{\SFR}{\Msunyr})}}}
\newcommand{\lMstar}{\ensuremath{{\log(\frac{\Mstar}{\Msun})}}}

\newcommand{\dd}{\text{d}}
\newcommand{\deriv}[2]{\frac{\dd{#1}}{\dd{#2}}}
\newcommand{\pp}{\ensuremath{\deriv{ \,P(\,\lsSFR\,|\,\Mstar) }{ \,\lsSFR }}}

\newcommand{\newchange}[1]{#1}
\newcommand{\nnewchange}[1]{#1}

\emergencystretch=\maxdimen
\title[The Main Sequence might be the tip of the iceberg.]{A single galaxy population? Statistical evidence that the Star-Forming Main Sequence might be the tip of the iceberg.}

\author[P. Corcho-Caballero et al.]{
P. Corcho-Caballero$^{1}$\thanks{E-mail: pablo.corcho@uam.es},
Y. Ascasibar$^{1}$,
\'A.R. L\'opez-S\'anchez$^{2,3,4,5}$
\\
$^{1}$Universidad Aut\'onoma de Madrid, Departamento de F\'isica Te\'orica, 28049, Cantoblanco, Madrid, Spain.\\
$^2$Australian Astronomical Optics, Macquarie University, 105 Delhi Rd, North Ryde, NSW 2113, Australia\\
$^3$Department of Physics and Astronomy, Macquarie University, NSW 2109, Australia\\
$^4$Macquarie University Research Centre for Astronomy, Astrophysics \& Astrophotonics, Sydney, NSW 2109, Australia\\
$^5$ARC Centre of Excellence for All Sky Astrophysics in 3 Dimensions (ASTRO-3D), Australia\\
}

\date{Accepted 2020 September 11. Received 2020 September 09; in original form 2019 December 12}

\pubyear{2020}

\begin{document}
\label{firstpage}
\pagerange{\pageref{firstpage}--\pageref{lastpage}}
\maketitle

\begin{abstract}
According to their specific star formation rate (sSFR), galaxies are often divided into `star-forming' and `passive' populations.
It is argued that the former define a narrow `Main Sequence of Star-Forming Galaxies' (MSSF) of the form $\sSFR(\Mstar)$, whereas `passive' galaxies feature negligible levels of star formation activity.
Here we use data from the Sloan Digital Sky Survey and the Galaxy and Mass Assembly survey at $z<0.1$ to constrain the conditional probability of the specific star formation rate at a given stellar mass.
We show that the whole population of galaxies in the local Universe is consistent with a simple probability distribution with only one maximum (roughly corresponding to the MSSF) and relatively shallow power-law tails that fully account for the `passive' population.
\newchange{
We compare the quality of the fits provided by such unimodal ansatz against those coming from a double log-normal fit (illustrating the bimodal paradigm), finding that both descriptions are roughly equally compatible with the current data.
In addition, we study the physical interpretation of the bidimensional distribution across the $M_*-\sSFR$ plane and discuss potential implications from a theoretical and observational point of view.
We also investigate correlations with metallicity, morphology and environment, highlighting the need to consider at least an additional parameter in order to fully specify the physical state of a galaxy.
}
\end{abstract}




\section{Introduction}

During the last 20 years large automatic surveys like the \textit{Sloan Digital Sky Survey} \citep[SDSS,][]{2000York, 2017Blanton} or the \textit{Galaxy And Mass Assembly} survey \citep[GAMA,][]{2011Driver, 2015Liske, 2018BaldryGAMA} have found that some fundamental parameters of galaxies (e.g. colors, stellar mass, $D4000$, angular momentum or SFR) display a bimodal distribution with two dense regions and a dip in between.

In particular, the distribution of galaxies along the colour-magnitude diagram \citep{2001Strateva, 2004Baldry, 2006Baldry, 2015Taylor} has often been interpreted in terms of two different galaxy populations: blue (`active') galaxies dwell at the so-called \textit{Blue Cloud} (BC), and they usually have younger stellar populations (i.e. recent star formation) than red (and `dead') galaxies which, in contrast, remain `passive' or `retired' (negligible star-forming activity) at the \textit{Red Sequence} (RS).
In general, low-mass galaxies tend to be located across the BC, whereas most massive systems can be found at the RS.
Morphology and gas content have also been shown to correlate with this trend \citep{2014Schawinski, 2017Eales, 2018Kelvin, 2018Bremer}; the BC is primarily composed of gas-rich, late-type galaxies, as spirals or irregulars, whereas those in the RS, typically early-type (elliptical or lenticular) galaxies, are thought to have negligible cold gas reservoirs \citep[][]{2008Kaviraj, 2016Tacchella, 2017Oemler, 2019SalvadorRusinol}.
The underdense space separating the BC and the RS is known as the \textit{Green Valley} (GV), and it is thought to represent the crossroads of galaxy evolution \citep{2007Wyder, 2007Martin, 2014Salim, 2014Schawinski, 2019Phillipps}. 
Basically, this region can be seen as \newchange{an intermediate stage} through which galaxies must transit before they reach the red sequence, and much effort has been devoted to unveil the underlying mechanism(s) responsible for driving galaxies from the BC to the RS, leaving an empty region in between \citep{2012Gonzalves, 2014Salim, 2019Quai, 2020Bluck}.

An important aspect to shed light on this question is whether the transition is triggered by a particular event that takes place at some precise moment in the history of a galaxy and shuts down star formation on a timescale much shorter than the age of the universe.
Since there is not a consistent convention throughout the literature, let us please stress that we will use the word \emph{quenching} exactly with this meaning, and the term \emph{ageing} to denote smooth variations of the star formation rate over scales comparable to the Hubble time.

The physical agents that could trigger quenching may be internal, such as gas outflows induced by supernova- or active galactic nucleus (AGN)-driven winds \citep{2005DiMatteo, 2005Springel, 2006Bower, 2011Kaviraj}, or external, like mergers with other galaxies \citep{2015DaviesLJM, 2017Weigel} or ram-pressure stripping in dense environments \citep{1972Gunn, 2008McCarthy, 2011McGee, 2019Daviesb}.
In this scenario, the galaxy population is indeed bimodal: star-forming (BC) galaxies are those that have never experienced a quenching event.
Quenched galaxies, on the other hand, become passive on a very short timescale, cross the GV in less than 1~Gyr, and then remain in the RS forever \citep[e.g.][]{2006Blanton, 2006Bundy, 2007Faber, 2010Peng}.
Less extreme variants, where star formation would decay more slowly, would be the \emph{strangulation} or \emph{starvation} of the gas reservoir \citep{1980Larson, 2002Bekki, 2008vandenBosch, 2015Peng, 2017Oemler, 2018Eales_strangulation} once a galaxy falls into a larger system, the inability to cool down once the dark matter halo reaches some mass threshold \citep{2003BirnboimDekel, 2006DekelBirnboim, 2006Cattaneo} or the stability of the cold gas disk against fragmentation (and thus, star formation) when the galaxy acquires a spheroidal-dominated morphology \citep{2009Martig}.
In this case, there are still two populations, but `quenched' galaxies would spend more time (a few Gyr) in the GV, and they would feature intermediate levels of star formation on their way to an asymptotically `passive' state on the RS.

Alternatively, it has also been argued that, for the vast majority of galaxies, star formation varies gradually at all times, from the BC to the RS, without any sharp transition \citep[see e.g.][]{2013Gladders, 2015Casado, 2015Abramson, 2016Abramson, 2017Eales, 2018Eales, 2018Bremer, 2019Phillipps}.
There may be variations of the star formation rate on short time scales, associated to events and/or processes internal or external to any given galaxy, but they may be regarded as random fluctuations around its mean star formation history, and they play a minor role on the evolution of the overall galaxy population.
The precise shape of the mean star formation history may also depend systematically on a number of factors, such as halo mass, concentration, angular momentum, large-scale environment, etc.  
\citep[e.g.][]{2016Erfanianfar, 2017GonzalezDelgado,2018Wang, 2019Popesso},
but there is not any quenching event to be singled out nor any clear distinction between the different stages of galactic evolution.
Here we will use the term \emph{ageing} to characterise those scenarios where, on average, most galaxies undergo secular evolution along cosmic time, slowly migrating from the BC to the RS as the specific star formation rate decreases progressively.
At any given time, different objects will be found at different evolutionary stages, but there is a single, continuous population, \newchange{even if the distribution of galaxy colours is clearly bimodal.
As pointed out by e.g. \citet{2017Eales, 2018Eales}, stellar populations much younger than about a Gyr are roughly equally blue, whereas much older populations are equally red, leading to two well-defined peaks in the probability distribution.
}


The distribution of galaxies on the plane defined by their stellar mass $M_*$ and their specific star formation rate $sSFR = SFR/M_*$ provides a snapshot of the \newchange{galaxy formation} process.
During the last decade many studies \citep[e.g.][]{2007Noeske, 2011McGee, 2011Elbaz, 2014Whitaker, 2015RenziniPeng, 2015Lee, 2016Cano, 2016Tacchella, 2017DuartePuertas, 2017Oemler, 2019Davies, 2019Caplar} have proposed that galaxies in the BC form a tight sequence in the $M_*-sSFR$ plane, which could even be just the projection of a more general relationship between stellar mass, star formation rate, and metallicity \citep[e.g.][]{Mannucci+2010, Lara-Lopez+10FP, Lara-LopezLSH13}.
For these BC galaxies, the relation can be roughly approximated as a power law with logarithmic slope close to linear and a normalisation that varies strongly as a function of redshift.
Detached from the so-called \emph{Main Sequence of Star-Forming galaxies} (MSSF), `retired' galaxies, roughly corresponding to the red sequence, are usually discriminated by imposing an arbitrary threshold in colour or sSFR.
It is only very recently \citep{2017Eales, 2018Eales} that a consistent description has been attempted in the framework of a more general `sequence' encompassing all kinds of galaxies.

The main objective of this work is to characterise the full statistical distribution of galaxies across the $\SFR$/$\sSFR$-$\Mstar$ plane, avoiding any selection based on colour cuts or activity thresholds, and test whether the data are compatible with a single \emph{ageing} population as an alternative to the bimodal scenario consisting of `active' and `quenched' galaxies.
More precisely, we focus on the conditional probability of the specific star formation rate at a given stellar mass, \pp, for galaxies in the Local Universe.
While one can always define a `sequence' of the form $\sSFR(\Mstar)$ in terms of the mode or the mean of this probability distribution, we highlight the importance of its tails in order to understand the diversity of star formation histories at fixed stellar mass \citep[e.g.][]{2015Abramson, 2016Abramson}.

The structure of this paper is as follows. Section~\ref{sec:observational_sample} describes the sample of observational measurements compiled from the literature upon which the present work is based.
The results of our analysis are presented in Sect.~\ref{sec:results}, while Sect.~\ref{sec:discussion} is devoted to their physical interpretation and the comparison with previous work.
Our main conclusions are briefly summarised in Sect.~\ref{sec:conclusions}.

For this work we use a $\Lambda$CDM cosmology with H$_0=70~\text{km}/\text{s}/\text{Mpc}$, $\Omega_{\Lambda}=0.7$ and $\Omega_{m}=0.3$.

\begin{figure*}
	\centering
	\includegraphics[width=.46\textwidth]{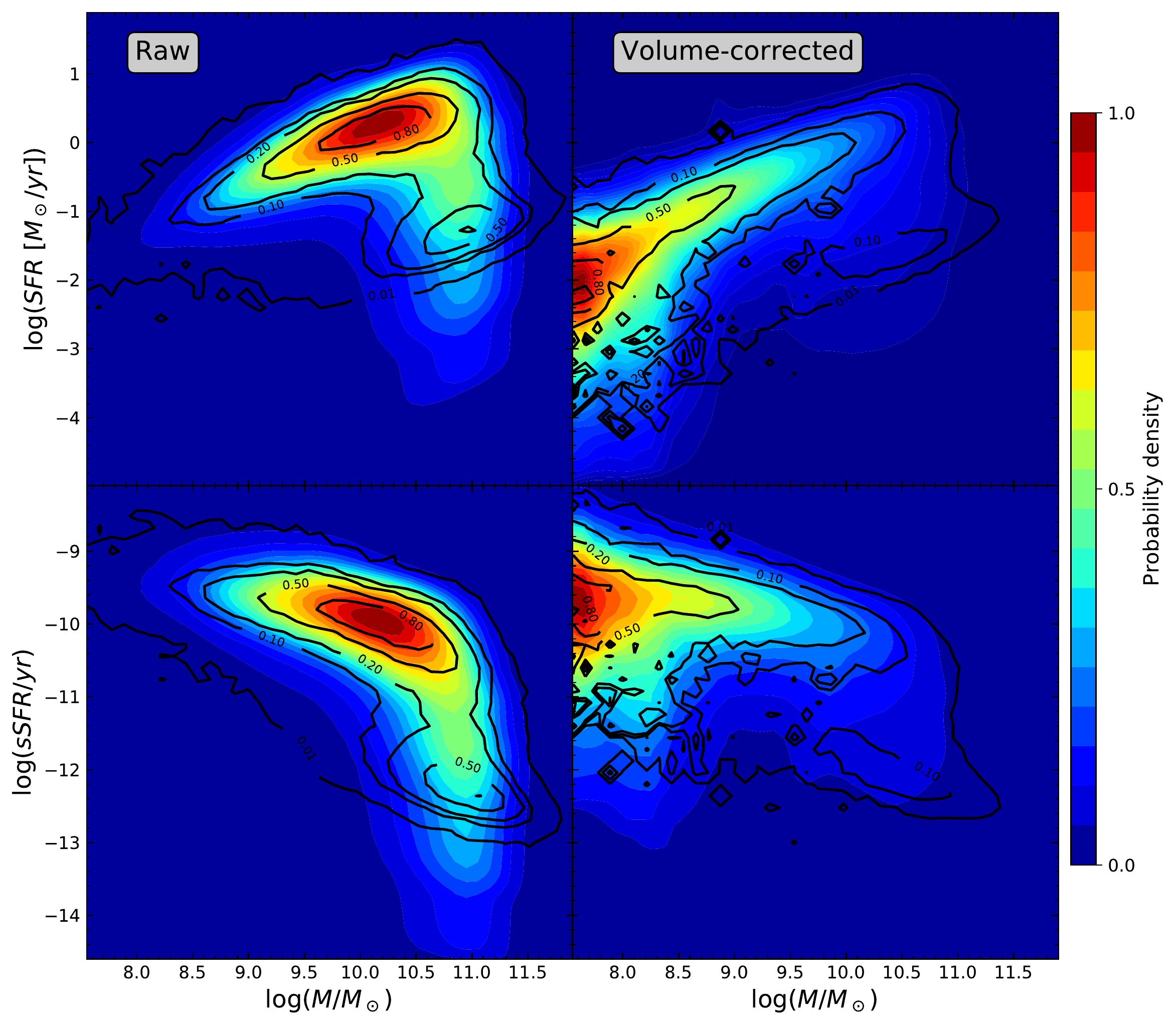}
	\includegraphics[width=.46\textwidth]{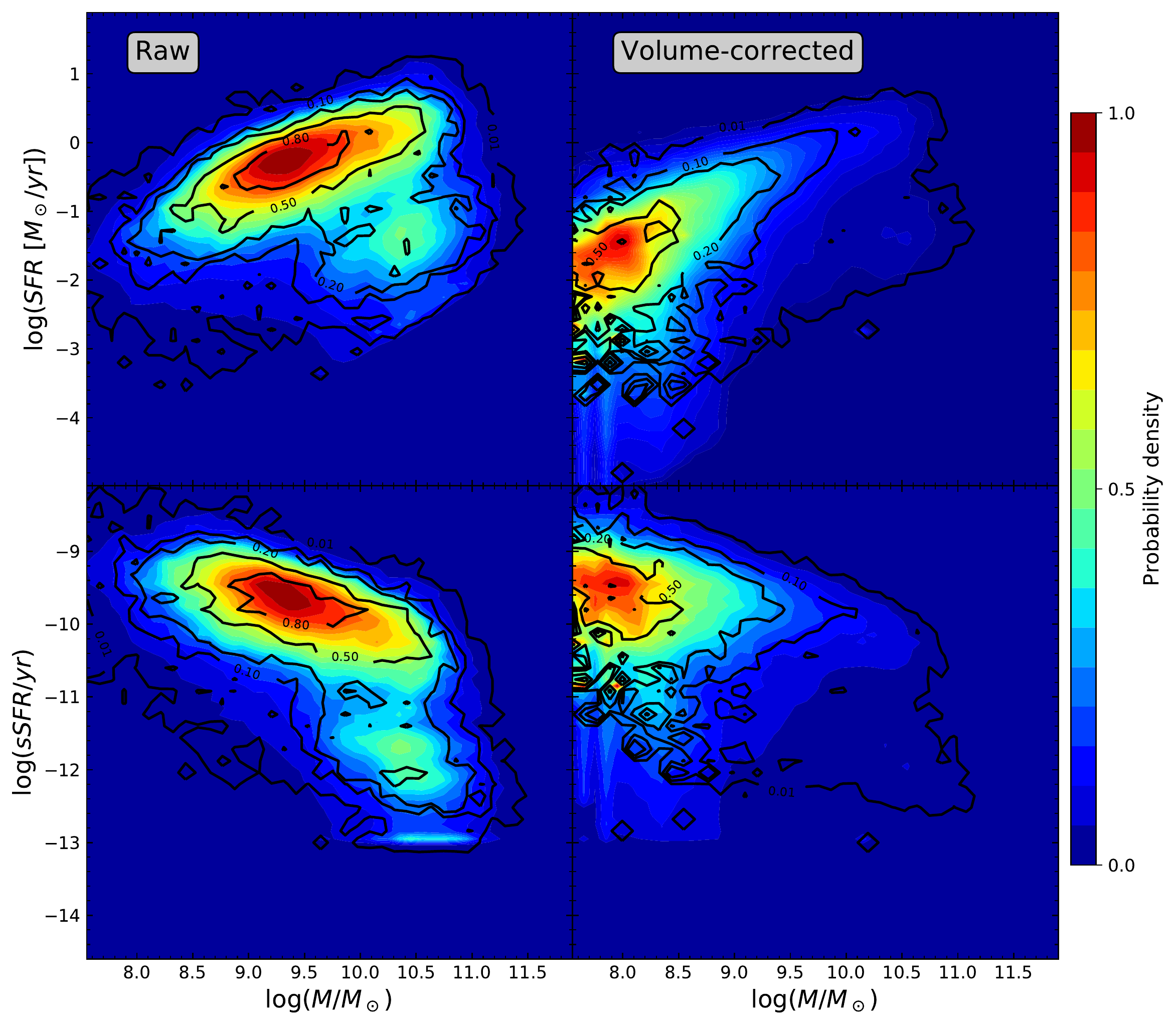}
	\caption
	{
		Bi-dimensional probability distribution of SDSS (left panels) and GAMA galaxies (right panels) in the $M_*$-SFR (\textit{top rows}) and $M_*$-SFR plane (\textit{bottom rows}).
		Black contours correspond to histograms computed from the estimated median values of $M_*$ and SFR/sSFR, and they enclose 1, 10, 20, 50 and 80 percent of the total sample. Coloured maps correspond to the probabilistic distribution computed as the sum of each individual pdf, where each color step increases in 5 percent the sample completeness. Left column histograms have been computed using non-corrected data whereas right column histograms have been volume corrected.
	}
	\label{fig:sdss_histograms}
\end{figure*}

\section{Observational sample}
\label{sec:observational_sample}

The primary galaxy sample we use in this work has been extracted from the photometric and spectroscopic galaxy surveys provided by the SDSS Data Release 16 \citep{2017Blanton}.
We consider the following selection criteria:

\begin{itemize}
    \item $0.001<z<0.1$,
    \item $m_{u,\text{petro}}<19.0, \hspace{0.4cm} ~m_{g,\text{petro}}<19.0, \hspace{0.4cm} ~m_{r,\text{petro}}<17.7$.
    \item CLASS = GALAXY
\end{itemize}
The first criterion reduces the galaxy sample to objects of the local universe, with lookback times below $\sim1~\text{Gyr}$, spatially restricted to a solid angle with inner and outer radius $D\sim4~\text{Mpc}$ and $D\sim450~\text{Mpc}$, respectively. 
The upper limits in magnitude ensure an homogeneous statistical selection and a minimum signal-to-noise ratio. 
Last criterion ensures that our sample is uniquely composed by galaxies, avoiding QSO's or faint stars.

Total stellar mass, SFR, and oxygen abundance, $12+\log(\text{O}/\text{H})$, measurements for each galaxy in the sample are obtained from the MPA-JHU spectroscopic catalogue\footnote{\textit{galSpecExtra} table from the value added catalogs.} (a detailed explanation of the methodology can be found at \citet{2003Kauffmann, 2004Brinchmann, 2004Tremonti}). \newchange{Stellar masses were computed by means of the $ugriz$ photometry, with small corrections performed using the fiber spectra.
SFRs were derived by combining emission line measurements for galaxies with enough S/N, and model fits to the integrated photometry for AGN and galaxies with weak emission lines.}
With this, our dataset comprises the spectroscopic redshift, $\{ugr\}$ Petrosian magnitudes, 50 and 90 Petrosian-flux radius and $\{2.5, 16, 50, 84, 97.5\}$ percentiles of stellar masses, SFR's and oxygen abundances of 162,279 galaxies\footnote{Note that galaxies with no emission lines will not have oxygen abundance measurements.}.

Complementary, we select another sample of galaxies from the GAMA Data Release 3 \citep{2018BaldryGAMA} with the same redshift cuts but extending the flux limit to $m_{r,\text{petro}}<19.8$, $m_{u,g,\text{petro}}<23.0$.
As with the SDSS sample, the stellar masses and SFR's percentiles are taken from the MagPhys data-product table, which are computed by fitting panchromatic SED's (21 photometric bands) using MAGPHYS code \citep{2008da_Cunha} as explained in \cite{2016Driver}\footnote{\textit{MagPhys} (Version 06) table from GAMA schema browser.}, comprising a total of 15,625 galaxies. 

We have corrected our data for the effect that the cosmological redshift produces on the spectral energy distribution. For this, we
apply the K-correction by means of the public algorithm \emph{kcor} provided by \cite{2010Chilingarian}\footnote{\url{http://kcor.sai.msu.ru/}}. Nevertheless,
due to the proximity of all elements in the sample, the net effect on each photometric band in general introduces differences smaller than 0.1~magnitudes in the $r$-band, less than 0.15~magnitudes for the $g$-band and smaller than 0.30~magnitudes in the $u$-band. 

We applied the 1/V$_{\text{max}}$ \citep{1968Schmidt} correction method to estimate the number density of galaxies. 
The maximum comoving distance at which a certain galaxy with an absolute magnitude $M_i$ would be included in our sample is given by the expression:
\begin{equation}
    \frac{D_{\text{max}}}{\text{Mpc}} = 10^{-0.2(M_i -m_{i,\text{lim}}+25+K_i)},
\end{equation}
where $m_{i,\text{lim}}$ represents our limiting apparent magnitude in each photometric band. Then, the maximum comoving volume is obtained applying
\begin{equation}
    V_{\text{max}} = \frac{\Omega}{3}\left(D_{\text{max}}^3 -D_{\text{min}}^3\right),
\end{equation}
where $\Omega$ is the solid angle covered by the survey $(\text{SDSS}~1,317~\text{deg}^2,~\text{GAMA}~180~\text{deg}^2)$\footnote{\url{https://www.sdss.org/dr16/scope},\\\url{http://www.gama-survey.org/dr3}}, $D_{\text{max}}$ is the minimum comoving distance obtained over the three photometric bands, and $D_{\text{min}}$ corresponds to the luminous distance at $z=0.001$.

\section{Results}
\label{sec:results}

\subsection{Revisiting the \Mstar-\sSFR\ plane}

We have explored the SFR/sSFR-$M_*$ plane by means of the extensive and widely used data-product sample provided by the SDSS collaboration and the multiwavelength analysis provided by the GAMA collaboration.
An important difference with respect to previous work is that we have accounted for measurement uncertainties by computing the probability density function (PDF, $\rho_i$) of each galaxy in the sample from the cumulative distribution function (CDF, $P_i$):
\begin{equation}
    \rho_i(\,\lMstar,\,\lsSFR\,) = \deriv{P_i(\lsSFR)}{\lsSFR} \deriv{P_i(\lMstar)}{\lMstar},
\end{equation}
where $P_i$ is reconstructed by linearly interpolating the published percentiles.
Then, the total probability distribution of the overall population is nothing but the sum of all individual pdf's, normalized according to the $V_\text{max}$ formalism:
\begin{equation}
\deriv{^2 P}{\,\lMstar\,\dd\lsSFR}
= \frac{ \sum_i^N \rho_i / V_{\text{max}, i} }{ \sum_i^N 1 / V_{\text{max}, i} }.
\end{equation}

The resulting galaxy distributions in the $\Mstar-\SFR$ and \mbox{$\Mstar-\sSFR$} planes are represented in the top and bottom panels of Fig.~\ref{fig:sdss_histograms}, respectively.
To illustrate the effect of the Malmquist bias, plots on the left column of each panel are computed without applying the $V_\text{max}$ correction.
In addition to our results, represented by the colour maps (where each step contains 5\% of the total probability), for the sake of comparison we also display as black contours the histograms computed by just using the median estimates of $\lMstar$ and $\lSFR$ (top) and $\lSFR$ (bottom) of SDSS galaxies, assuming that $\rho_i$ is a delta function (i.e. ignoring uncertainties).

Both observational effects are important.
As it is well known, volume correction shifts the peak of the distribution to the low-mass end, where there is a high concentration of dwarf galaxies\footnote{For simplicity, we  address as `dwarf' galaxy any system whose stellar mass is below $10^9 \text{M}_\odot$.} around $\sSFR \sim 10^{-9.5}$~yr$^{-1}$ (i.e. forming stars at a rate that is significantly above their past star formation history), as well as a significant fraction of `passive' systems with much lower sSFRs.
These objects are sometimes interpreted as quenched satellite galaxies \citep[e.g.][]{2015RenziniPeng}, but we do not observe any bimodality in the distribution, only a continuous tail towards low star-formation activity, let alone a signature of rapid quenching.

\nnewchange{It is important to notice that we define quenched galaxies as being clearly detached from the star-forming systems.
At low stellar mass, stochastic star formation would broaden both the MS and the statistical distribution of `passive' galaxies, that could in principle undergo subsequent bursts.
Our approach is only sensitive to the overall probability distribution and cannot discriminate between smooth and stochastic star formation histories, but in either case we do not find evidence that low-mass galaxies can be divided in two different populations.
In order to be viable, the bimodal scenario of `star-forming' and `quenched' galaxies would require that their statistical distributions overlap, and at least one of them has significant tails.
Completely passive galaxies should be absent from our sample due to selection effects, and the MS should contain a mixture of `star-forming' and (physically distinct) `rejuvenated' galaxies.
We argue that our current data may be interpreted in terms of stellar feedback processes (dominant mechanism in field galaxies), which would create a continuous distribution of sSFRs, not being able of completely suffocating star formation but regulating it, either smoothly or stochastically, leading to bursty episodes of star formation. However, this does not preclude other processes (e.g. strangulation in dense environments) for being responsible of a dwarf passive population below our detection threshold.}
Therefore, even larger galaxy samples are needed to establish more robust conclusions in this regime.
To some extent, the same is true also in the mass range between $10^9$ and $10^{10}$~\Msun, where it would be interesting to confirm more confidently whether the width of the probability distribution does actually reach a minimum, as suggested by both SDSS and GAMA data.

For massive galaxies, volume corrections become negligible, but their number is limited by the survey volume.
Moreover, they tend to populate the lower sSFR part of the diagram, and thus their star formation activity is much more difficult to estimate accurately.
Measurement uncertainties are particularly relevant in the SDSS, where the peak associated to `passive' galaxies in the distribution of median values (black contours) is completely washed out when errors are taken into account.
The uncertainties are much smaller in GAMA data, most likely because the ancillary observations at other wavelengths help breaking the degeneracy between stellar age, metallicity, and dust extinction.
Here we do observe that massive systems do indeed seem to concentrate around $\sSFR \sim 10^{-12.5}$~yr$^{-1}$, and that there are hints of a certain bimodality in the mass range between $10^{10}$ and $10^{11}~\Msun$.
The main potential issues arise in this case from the reduced number of galaxies ($\sim 10$~times smaller than the SDSS sample), and the apparent cut-off\footnote{The upper limit of the stellar mass at $\Mstar = 10^{11.5}~\Msun$ might also play a role on this.} at $\sSFR = 10^{-13}$~yr$^{-1}$, most likely associated to the priors assumed in the data analysis, that constraints all measurements to be above this threshold.
From these evidence alone, we would conclude that the question of whether there is a minimum \sSFR\ in the local universe, especially at high masses, remains open.
Accurately measuring such low levels of star formation activity is extremely challenging, but it is of the utmost importance in order to settle this issue.

\begin{figure*}
	\centering
	\includegraphics[width=0.49\linewidth]{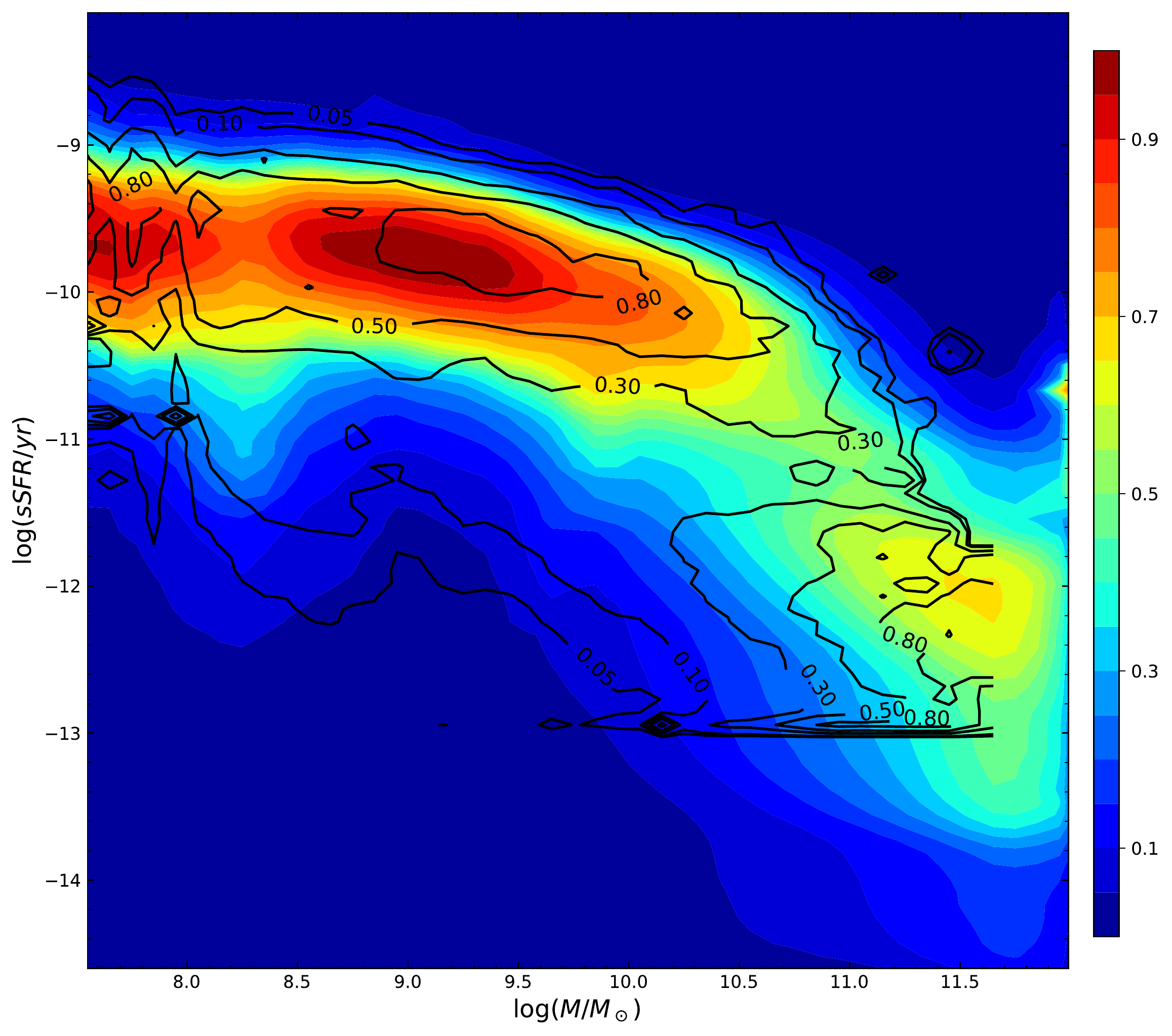}
	\includegraphics[width=0.49\linewidth]{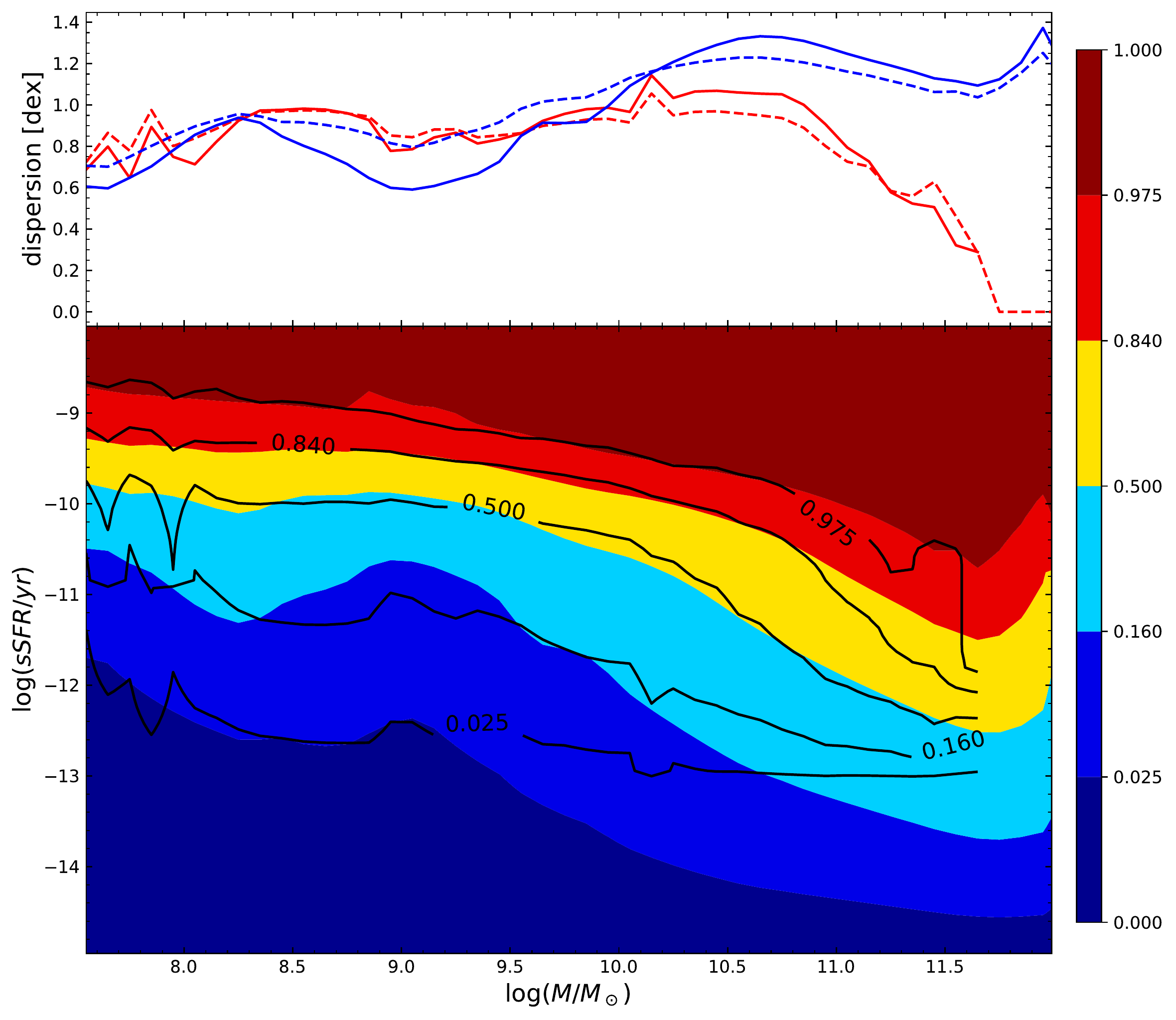}
	\caption{Left panel shows the volume-corrected conditional probability $\Mstar-\sSFR$ plane. Colored contours correspond to the distribution derived using SDSS data while black contours were computed with GAMA data. Each color step increases in 5 percent the sample completeness. 
	Bottom-right panel represents the volume-corrected conditional cumulative probability distribution of the sSFR-$M_*$ plane for SDSS (coloured fringes) and GAMA (black lines).
    Top-right panel represents the difference between the 84 and 16 percentiles divided by 2 (solid lines) and the second moment of the $\lsSFR$ distribution (dashed lines) as function of stellar mass: blue for the SDSS and red GAMA.
}
	\label{fig:conditional}
\end{figure*}

\subsection{Conditional probability}

\begin{figure*}
	\includegraphics[width=\linewidth]{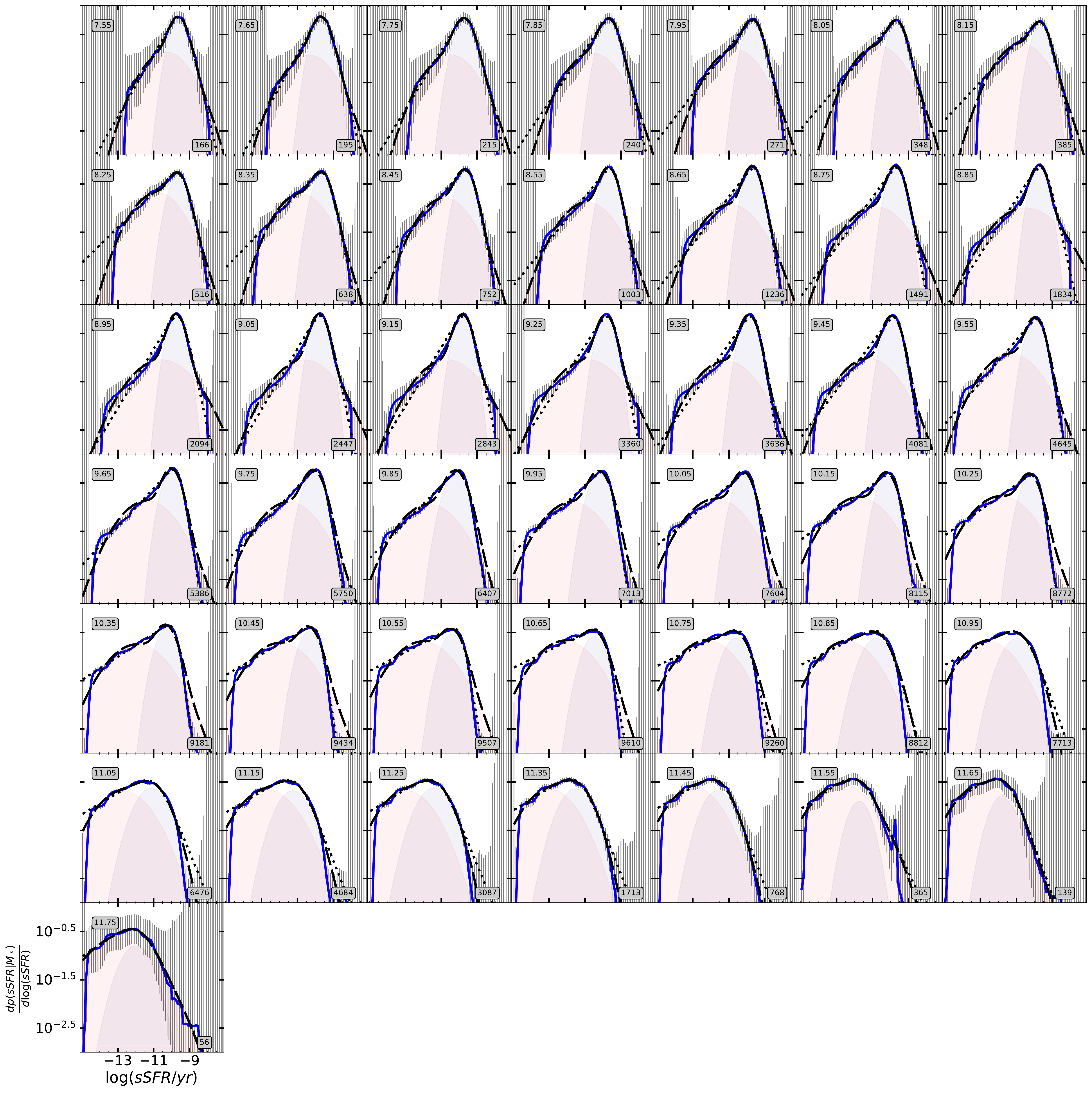}
	\caption{
		Probability distribution of $\lsSFR$ in 43 mass bins. The corresponding mid-bin mass, $\sim\log(\Mstar/\Msun) \pm 0.05$, is indicated at the upper left corner of each panel as well as the number of galaxies within each bin at the lower right corner.
		The SDSS data are shown in blue, with error bars in grey; black short-dashed lines denote the ansatz $\hat p_{\text{uni}}$ \eqref{eq:ansatz}; black long-dashed lines denote the ansatz $\hat p_{\text{bim}}$ \eqref{eq:lognorm} and blue/red areas illustrates the individual `active' and `passive' populations.
	}
	\label{fig:sdss_fits}
\end{figure*}

The conditional probability
\begin{equation}
p \equiv \pp,
\label{eq:distribution}
\end{equation}
plotted on the left panel in Fig.~\ref{fig:conditional} provides a complementary view of the data.
Once again, we find fair agreement between the SDSS (colours) and the GAMA (black contours) data in the low-mass, high-sSFR range, with a maximum conditional probability around \mbox{$\sim3 \times 10^{-1}$~Gyr$^{-1}$} for all masses below $\sim 3 \times 10^{10}$~\Msun.
This result has usually been interpreted as a tight `star forming sequence'.
At the high mass end, the mode of the conditional probability moves towards lower \sSFR, but its precise shape is difficult to measure, and it is extremely sensitive to the data analysis.
When uncertainties are taken into account in the SDSS data, the maximum probability decreases smoothly with stellar mass, with no robust indications of bimodality \nnewchange{for any value of $M_*$ (see Figure~\ref{fig:sdss_fits} below for a more detailed representation)}.
However, the GAMA data feature smaller error bars, and a narrower mass range (between $\sim 10^{10}$ and $10^{11}$~\Msun) where the conditional probability of the specific star-formation rate seems to display two peaks.

From these data, we consider that there is only marginal evidence for a bimodal conditional probability in this narrow mass range.
Confirming this result is of the utmost importance: if the conditional probability presented a single maximum for all stellar masses, the galaxy population could be described in terms of a single `sequence' with a well-defined mode; if were indeed bimodal, such a description would not be valid (although it would not necessarily imply a physical bimodality in the galaxy population; see discussion).
Here we will assume that the conditional probability has always a single peak, but a larger sample size is required to increase the statistical significance in the intermediate mass regime, as well as a thorough discussion of the systematic methodological uncertainties associated to the derivation of the SFR.

In any case, \nnewchange{we would like to argue} that this `sequence' can not be understood as a tight relation of the form $\sSFR(\Mstar)$, \nnewchange{given the relatively high scatter, reaching almost one order of magnitude. Moreover}, both data sets show that the conditional probability has prominent tails that extend far away from the mode for any given stellar mass.
In order to obtain a more robust statistical characterisation of the probability distribution, right panel of Figure~\ref{fig:conditional} represents the dispersion of the \sSFR\ as as function of stellar mass (top panel), as well as the $2.5$, $16$, $50$, $84$, and $97.5$ percentiles.

At low stellar mass, both samples yield remarkably consistent results, with a clear correlation between the stellar mass and the present activity of galaxies.
All percentiles follow a decreasing path from the most active dwarf galaxies towards a much milder star formation activity in massive systems.
We advocate that the smoothness of the variation of the percentiles (which can be more robustly derived than the mode) favours the interpretation in terms of a single galaxy population ageing steadily.
Although this is consistent with the idea of a galaxy sequence as proposed by numerous authors, we point out that the dispersion is, \nnewchange{in our view}, far too large to consider it a tight relation between \Mstar\ and \sSFR.
The dispersion of the distribution, quantified in terms of the second moment or the difference between percentiles, increases slightly as a function of \Mstar, but it is always of the order of a factor of $\sim 6-10$.

In agreement with previous works \citep[e.g.][]{2015Guo, 2015Willett, 2019Davies}, we find hints that the dispersion might reach a minimum around $\sim 10^9~\Msun$, specially for the SDSS data, with a steeper tail towards low \sSFR.
The origin of this 'U' shape could be attributed to uncertainties in the measurements, survey selection effects or physical processes, but studying its nature is out of the scope of the present work.
At the high mass end, it seems evident that the low number statistics and the artificial \sSFR\ threshold in the GAMA data preclude a meaningful comparison.

In general, our values of the dispersion are higher than those discussed in the literature, specially at the edges of the distribution, because we do not apply any selection cut.
It is our main aim to highlight the importance of quantifying the conditional probability distribution of the \sSFR\ as a function of stellar mass, not only the mean/mode downsizing trend (the so-called `main sequence'), but also the dispersion and the asymmetric long tails that are necessary in order to fully characterise the intrinsically bidimensional distribution in the $\Mstar-\sSFR$ plane.

\begin{figure*}
	\includegraphics[width=\linewidth]{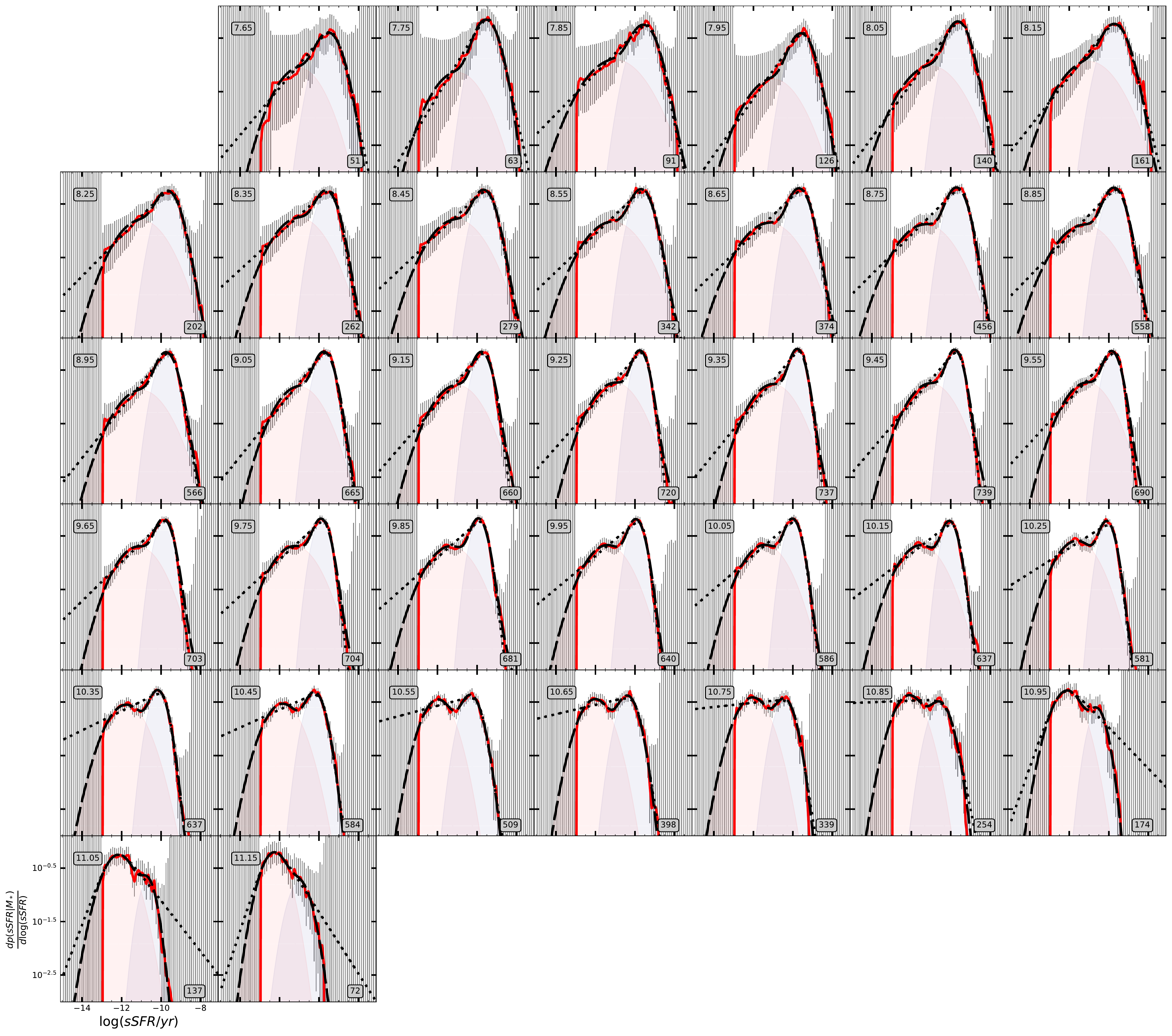}
	\caption{
		Probability distribution of $\lsSFR$ in 36 mass bins. The corresponding mid-bin mass, $\sim \log(\Mstar/\Msun) \pm 0.05$, is indicated at the upper left corner of each panel as well as the number of galaxies within each bin at the lower right corner.
		The GAMA data are shown in red, with error bars in grey; black short-dashed lines denote the ansatz $\hat p_{\text{uni}}$ \eqref{eq:ansatz}; black long-dashed lines denote the ansatz $\hat p_{\text{bim}}$ \eqref{eq:lognorm} and blue/red areas illustrates the individual `active' and `passive' populations.
	}
	\label{fig:gama_fits}
\end{figure*}

\subsection{Analytical characterisation}
\label{sec:Analytical characterisation}

The distribution of galaxies in the Local Universe in the $\sSFR-M_*$ plane may be interpreted as some sort of evolutionary pathway seen from a snapshot at present time, but the underlying physical processes that regulate star formation are diverse, and a galaxy of a given stellar mass may display a wide variety of possible evolutionary histories.
Rather than considering present states as a discrete set (e.g. 'main sequence' and 'quenched'/'quiescent' galaxies) based on arbitrary thresholds, we advocate for a continuous bi-dimensional distribution, given by the conditional probability~\eqref{eq:distribution}.

\newchange{
As an empirical estimate, Fig.~\ref{fig:sdss_fits} and \ref{fig:gama_fits} plot the probability distribution of \lsSFR\ corresponding to SDSS and GAMA (blue and red solid lines, respectively), in mass bins increasing in logarithmic steps of 0.1~dex with at least 50 galaxies per bin.
The error bars, shown in grey, were estimated as
\begin{equation}
\label{eq:hist_err}
    \sigma (\log(p)) = \frac{1}{\ln{10}\sqrt{N_i \rho_i \Delta\lsSFR}}
\end{equation}
where $N_i$ is the number of galaxies within that bin.
}

\newchange{
Once again, hints of bimodality in the intermediate-mass range $10<\lMstar<11$ can be observed in the GAMA data.
If this feature were indeed real (and SDSS measurements had buried it beneath the uncertainty level), a two-component fitting function would be necessary. 
On the other hand, it is also obvious from this plot that both tails of the distribution are better represented by power laws with different logarithmic exponents than a single Gaussian function.
To settle this issue, we have fitted two different models to the data and compared their efficiency: a bimodal approach and a unimodal distribution with power law tails.
}

\newchange{
Typically, the bimodal population of galaxies is represented as the sum of two log-normal distributions}
\begin{equation}
\begin{split}
\label{eq:lognorm}
\hat p_{\text{bim}}(&\sSFR, \Mstar | f_p, \mu_p, \sigma_p, \mu_a, \sigma_a) = \\
& \frac{f_p}{\sigma_p\sqrt{2\pi}} e^{-\frac{(\lsSFR-\mu_p)^2}{2\sigma_p^2}}
+ \frac{1-f_p}{\sigma_a\sqrt{2\pi}} e^{-\frac{(\lsSFR-\mu_a)^2}{2\sigma_a^2}}
\end{split}
\end{equation}
\newchange{where the subindices $a$ and $p$ stand for `active' and `passive', whereas $f_p$ denotes the `passive' fraction. 
The best fit for each data-set is represented as a black long-dashed line. 
In addition, `passive' and `active' populations are illustrated by reddish and blueish colored areas respectively, depicting their preponderance in each bin.
During the \nnewchange{fit $-14<\mu_p<-11$ and $-11<\mu_a<-8$ have been imposed}, aiming at a sensible interpretation in terms of two independent galaxy distributions.}

\newchange{Alternatively, in this work we propose a new and simple ansatz}
\begin{equation}
\label{eq:ansatz}
\hat p_{\text{uni}}(\sSFR, \Mstar | \alpha, \beta, \sSFR_0) \propto
 \frac{\left(\sSFR/\sSFR_0\right)^\alpha}{1+\left(\sSFR/\sSFR_0\right)^{(\alpha+\beta)}},
\end{equation}
where $\alpha$ and $\beta$ represent the asymptotic logarithmic slopes for low and high $sSFR$ values, respectively, and $\sSFR_0$ roughly corresponds to the transition between both regimes.
We consider that, for our present data set, this choice provides a reasonable compromise between accuracy and number of free parameters. \newchange{These fits are represented by black short-dashed lines.}

\newchange{In both figures we can see the fair agreement between the two descriptions and the observed distribution of galaxies, although the SDSS data appear to be slightly better fitted by our unimodal ansatz, while the GAMA data seem to be better represented with the bimodal description, especially in the mass range $9.5<\lMstar<11$.
An improved statistical sampling of the extremes of the distribution would be needed in order to assess which model, if any, provides a more accurate representation. In our current datasets, it is difficult to discriminate a sharp log-normal cutoff from a relatively shallow power law. At the passive end, methodological issues related to the measurement of the sSFR also play an important role.
}


\begin{figure}
	\includegraphics[width=\linewidth]{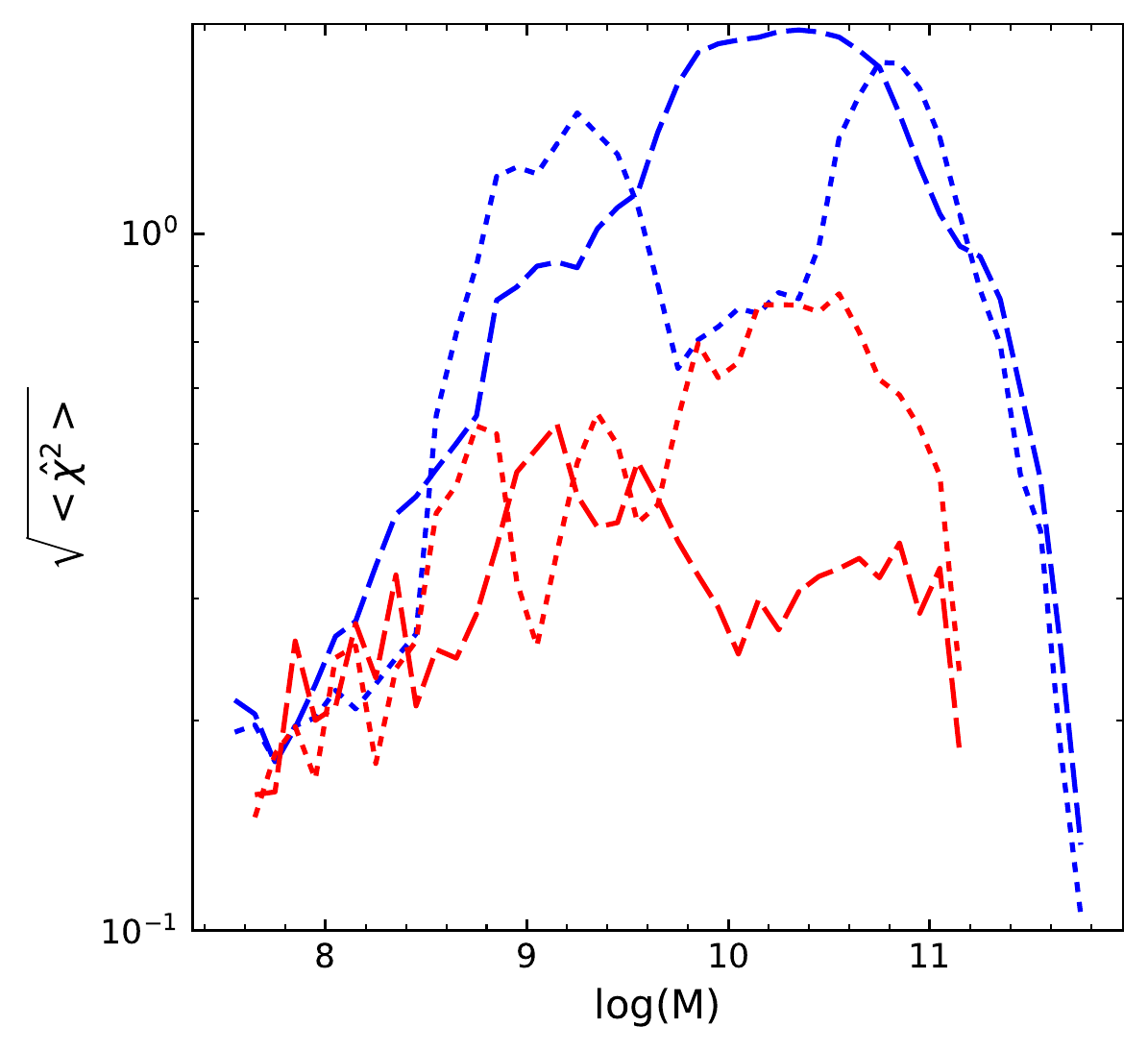}
	\caption{
		Values of $\sqrt{\chi^2}$ computed from fits shown in Figures \ref{fig:sdss_fits} and \ref{fig:gama_fits}, both SDSS \nnewchange{(blue)} and  GAMA \nnewchange{(red)} using the bimodal description \eqref{eq:lognorm} (long-dashed) and the unimodal distribution \eqref{eq:ansatz} (short-dashed). 
	}
	\label{fig:chi_square}
\end{figure}

\newchange{To quantify the performance of both approaches, Fig.~\ref{fig:chi_square} shows the mean values of $\sqrt{\chi^2}$ computed for each fit as
\begin{equation}
    <\hat{\chi}^2> = \frac{1}{N}\sum_i^N \frac{(\log(p(sSFR_i) - \log(\hat{p}(sSFR_i))^2}{\sigma(sSFR_i)^2}
\end{equation}
Blue and red colors illustrate the results with the SDSS and the GAMA data respectively.
Short-dashed lines denote our ansatz while long-dashed lines represent the bimodal fits. 
The relatively low values obtained, $\chi \la 1$, hint that the uncertainties obtained from expression~\eqref{eq:hist_err} may have been overestimated, but this does not preclude a meaningful comparison between both models and data sets in a relative sense.
In this figure we can clearly see that the region where bimodality is typically reported to appear (i.e. $9.5\lesssim \log(M/M\odot) \lesssim 11 $) is the most controversial regarding the model choice: SDSS data are better described by a single galaxy population while GAMA data indicate that two populations are more suitable.
Outside this range, both models yield very similar accuracy.
Consequently, we conclude that they are able to provide an equally feasible description of the data, and selecting one of them is a difficult (and somewhat subjective) task.}

\begin{figure*}
    \centering
    \includegraphics[width=0.8\linewidth]{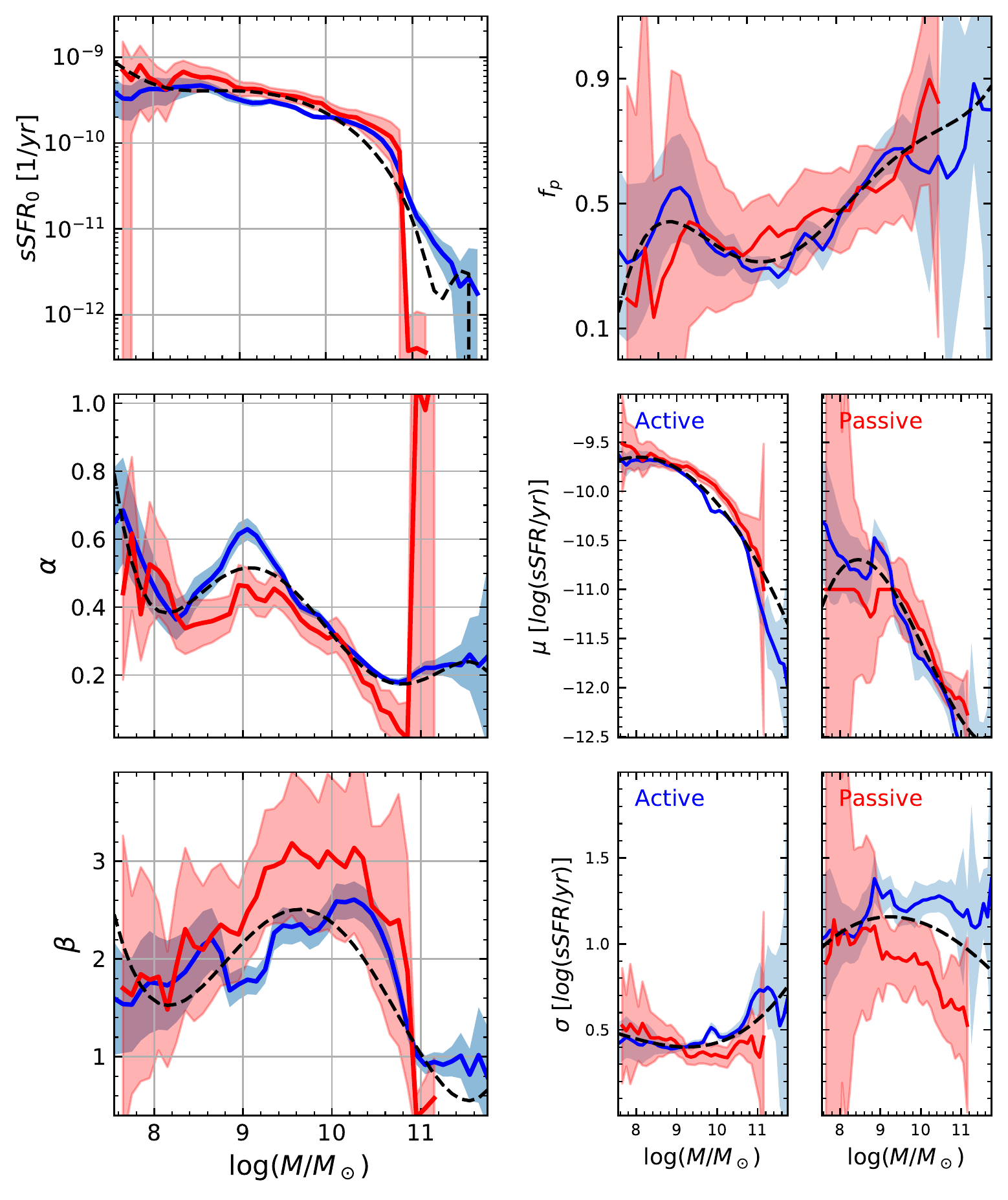}
    
    \caption{Fit's parameters mass-dependency. 
    \newchange{Left column panels contain the parameters used in the unimodal model and right column panels include the parameters used within the bimodal approach. Blue lines denote the results with SDSS data and red lines those obtained with GAMA data. Shady regions represents twice the standard deviation from the mean value. Black dashed lines denote the polynomial fits for each parameter}}

    \label{fig:fit_params}
\end{figure*}

\newchange{Figure~\ref{fig:fit_params} shows the evolution with mass of each parameter used in the models.
On the one hand, the left column plots correspond to the unimodal ansatz.}
The turnover $\sSFR_0$ is roughly constant at low stellar mass, and it steadily decreases beyond $\sim 10^9$~\Msun.
The $\beta$ parameter that controls the relative abundance of the most active galaxies (with respect to the `main' population near $\sSFR_0$) roughly displays a constant value $\beta\sim 2$ for $\lMstar\lesssim 10.3$, decaying gradually (i.e. flatter slope) at larger masses.
The low-activity tail captured by the logarithmic slope $\alpha$ exhibits a more complex behaviour. 
For dwarf galaxies, our individual fits seem to indicate, at face value, that the tail becomes shallower (i.e. a larger fraction of `passive' systems) up to $\lMstar\sim 8$, and then the slope increases, reaching a local maximum at $\lMstar\sim 9$, \newchange{possibly affected by Malmquist bias}.
For intermediate-mass galaxies, the best-fitting value of $\alpha$ descends again up to $\lMstar\sim 11$, and it finally remains flat beyond that limit.
\newchange{On the other hand, the right column panels display the variables used with the two log-normal model. As expected, the fraction of `passive' galaxies represented by $f_p$, becomes more relevant as mass increases, dominating completely the high-mass end of the distribution. 
The mean $sSFR$ of the `active' ($\mu_a$) and `passive' populations ($\mu_p$) describe a decreasing path similar as $sSFR_0$. In particular, $\mu_a$ might be interpreted as the location of the MSSF along the $\sSFR-\lMstar$ plane.
The shape of the standard deviation of `active' galaxies is roughly flat with typical values around $\sigma\sim 0.4-0.5$~dex, as previously reported on the literature \citep{2013Willet, 2015Guo, 2019Davies}.
The values of $\sigma_p$ are much higher, above $1 $~dex, and its precise shape remains unclear given the current accuracy of the data.}

\newchange{
We would like to highlight here that our unimodal ansatz has three parameters, compared to 5 for the double log-normal description.
Moreover, `active' galaxies in the bimodal scenario are always distributed in a narrower region than the `passive' population, which practically spreads across the whole range of $sSFR$ probed by the measurements, calling into questioning its physical interpretation, particularly for values beyond the MSSF.
One may avoid this problem by restricting the location and width of the distributions to ensure that both populations are meaningfully isolated, but in such case the passive tail could not be reproduced as accurately as it is done by the unimodal ansatz.
The systematic variation of $\mu_p$ with stellar mass is also not trivial to understand, and it may point to selection effects playing an increasing role at smaller stellar masses.
Once again, deeper observations will prove extremely helpful in exploring this issue.
}

\newchange{Finally, for the sake of reproducibility, we fitted each parameter ($\lambda$), with a polynomial 
\begin{equation}
    \lambda = \sum_i a_i \log(M/M_\odot)^i, 
\end{equation}
as shown in each panel of Fig.~\ref{fig:fit_params} with a black dashed line. The corresponding values for each coefficient are given at Table~\ref{tab:polynomial_fits}. }

\begin{table*}
    \centering
    \begin{tabular}{c|c|c|c|c|c|c}
        & $a_0$ & $a_1$ & $a_2$ & $a_3$ & $a_4$ & $a_5$ \\
        \hline
        $sSFR_0$ &  $-1.55\cdot10^{-11}$ & $7.80\cdot 10^{-10}$ & $-1.56\cdot10^{-8}$ & $1.55\cdot10^{-7}$&$-7.68\cdot10^{-7}$ &$1.51\cdot10^{-6}$\\
         $\alpha$& -0.032 & 1.496 & -29.40 & 287.04 & -1392.20 & 2684.11 \\
         $\beta$ & 0.175 & -6.83 & 99.25 & -634.23 & 1507.09 & -- \\
         $C$ & 0.032 & -1.55 & 29.94 & -288.16 & 1380.97 & -2630.35 \\
         $\mu_a$ & -0.123 & 1.99 & -17.66 & -- & -- & --\\ 
         $\mu_p$ & -0.034 & 1.704 & -33.61 & 329.05 & -1.599 & 3071.8\\ 
         $\sigma_a$ & 0.0068 & -0.1465 & 0.9324 & -1.1079 & -- & --\\ 
         $\mu_p$ & 0.021 & -0.652 & 6.765 & -21.95 & -- & --\\

    \end{tabular}
    \caption{Coefficient values of polynomial fits shown in Fig.~\ref{fig:fit_params}.}
    \label{tab:polynomial_fits}
\end{table*}



\begin{figure*}
	\includegraphics[width=.33\linewidth]{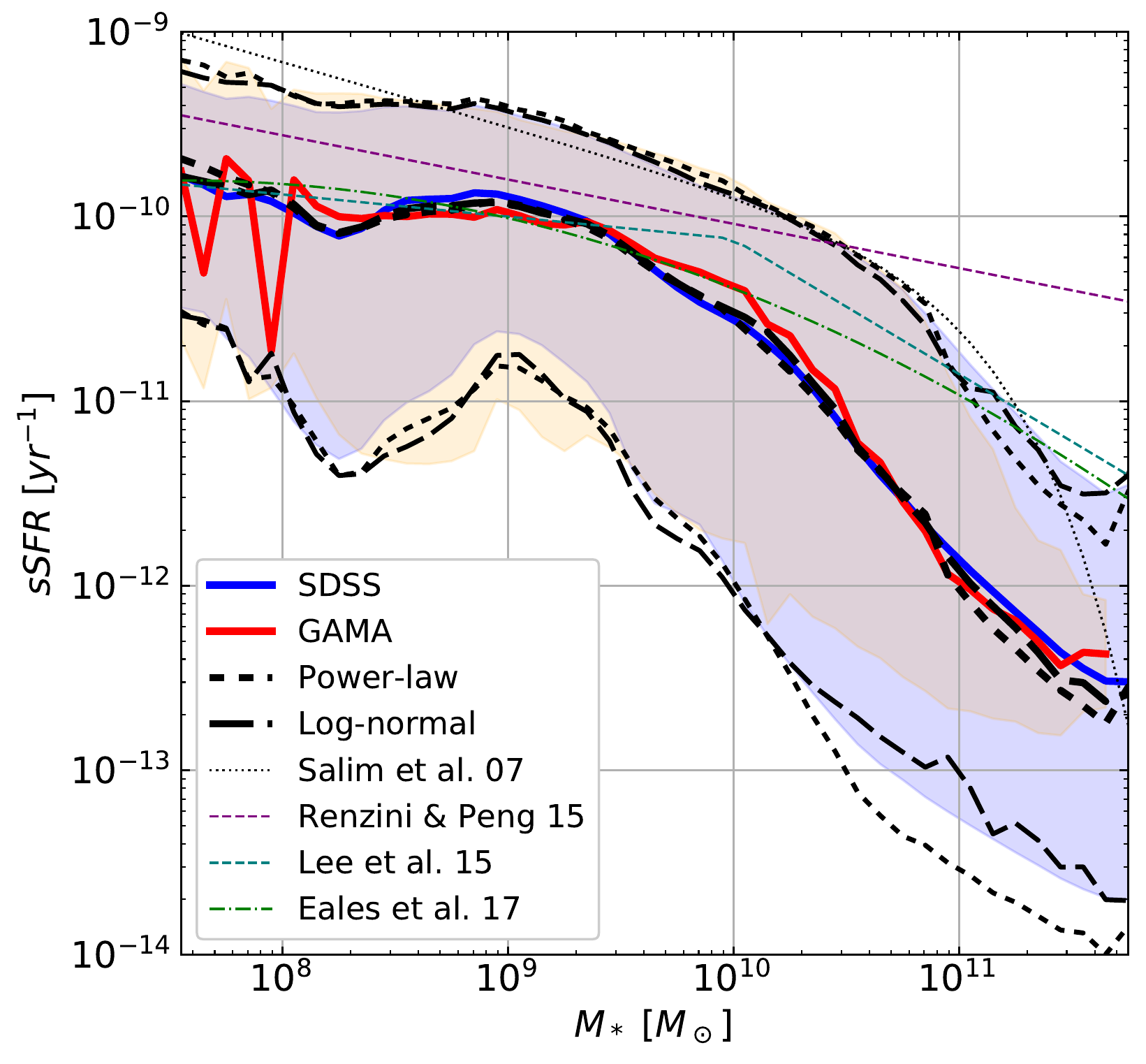}
	\includegraphics[width=.33\linewidth]{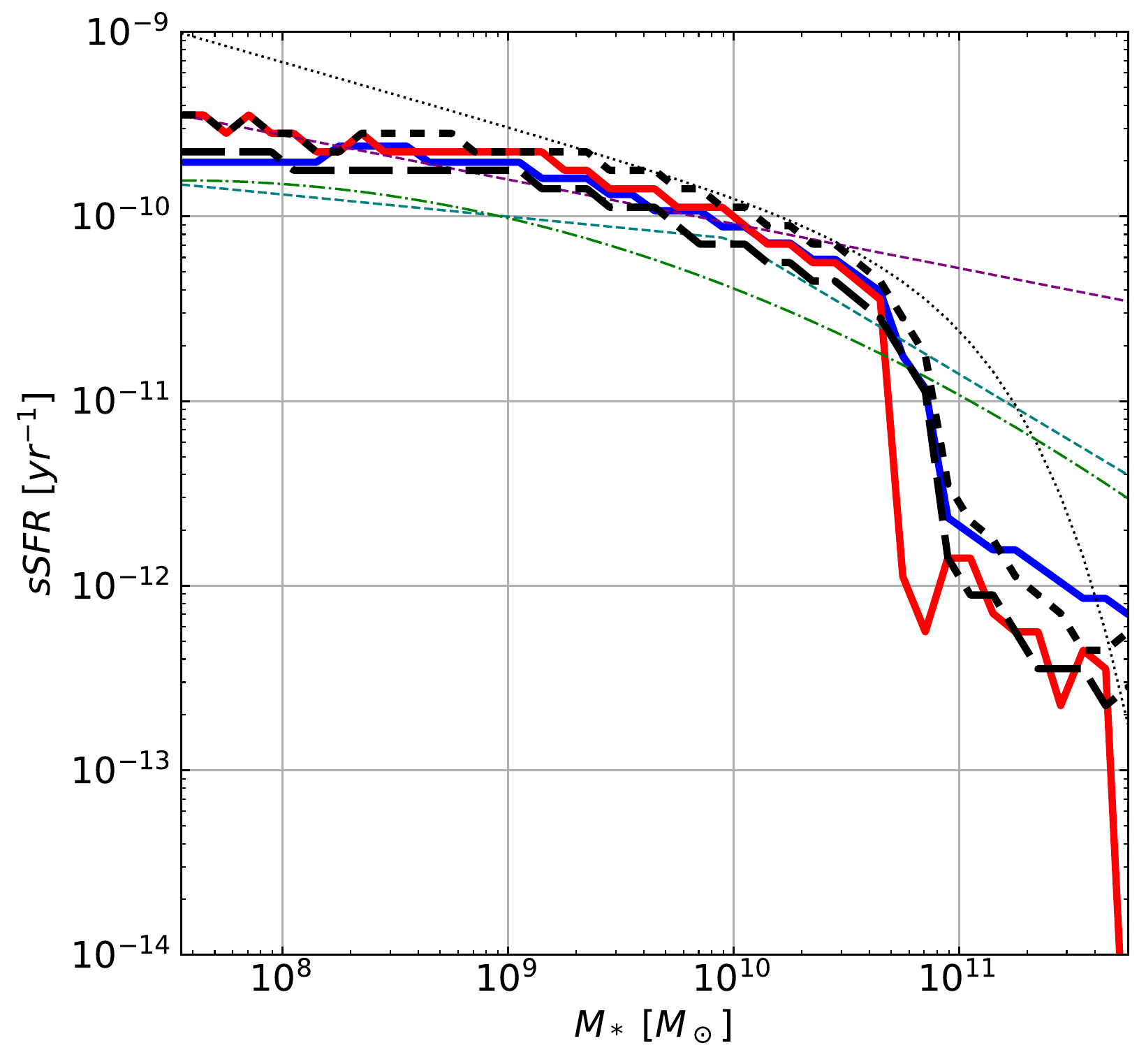}
	\includegraphics[width=.33\linewidth]{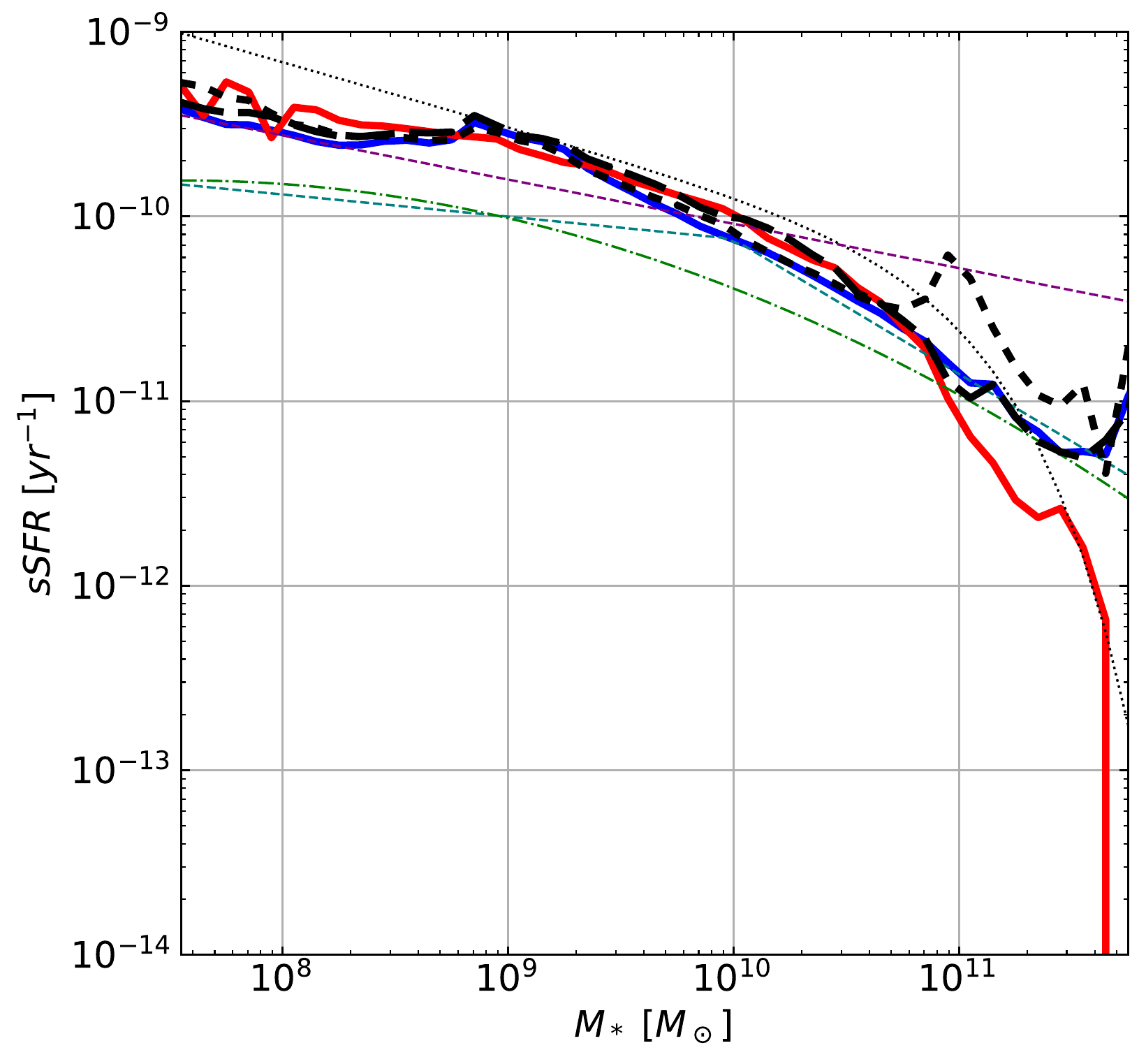}
	\caption{
		Conditional probability distribution of the sSFR as a function of stellar mass, $\deriv{p(\sSFR|\Mstar)}{\sSFR}$.
		Measurements based on SDSS and GAMA data are shown in blue and red colour, respectively; thick solid lines denote the median (left), mode (center) and mean (right) sSFR, while the shaded areas on the left panel illustrate the 16 and 84 percentiles. \newchange{Results from the unimodal distribution are shown by a short-dashed black line, whereas the bimodal results are displayed as long-dashed black lines.}
		Thin coloured lines correspond to different fits proposed in the literature (see legend).
	}
	\label{fig:sequence_fit}
	
\end{figure*}

\section{Discussion}
\label{sec:discussion}

From a physical point of view, we want to argue that:
\begin{enumerate}
\item Regardless of whether the distribution in the $M_*-sSFR$ plane is bimodal or not, we do not support the existence of two discrete states representing `star-forming' and `passive' systems, but a continuous evolution from the highest to the lowest \sSFR.
\item When all galaxies are included in the analysis, the conditional probability $p$ is not a narrow, symmetric log-normal that can be fully characterised in terms of a one-parameter sequence of the form $sSFR(M_*)$.
Although a correlation does indeed exist, the stellar mass alone does not specify the evolutionary state of a galaxy.
\end{enumerate}{}

\subsection{A single galaxy population}
\label{subsec:fits}

The distribution of galaxies in the $\Mstar-\sSFR$ plane provides an important constraint to galaxy formation and evolution models, and it has been widely studied in the literature.
To date, most efforts have focused on characterising the mean \citep[e.g.][]{2015Lee, 2017Eales}
\begin{equation}
	\langle sSFR \rangle(M_*) = \int_0^\infty sSFR \deriv{p(\sSFR|\Mstar)}{\sSFR}\ \dd\,\sSFR
\end{equation}
or the mode \citep[e.g.][]{2007Salim, 2015RenziniPeng} of the distribution, often excluding `passive' galaxies according to different criteria.

On this regard, Fig.~\ref{fig:sequence_fit} plots the percentiles, mode, and mean sSFR of the conditional probability distribution derived with the SDSS/GAMA data (blue and red solid lines, respectively), compared to our ansatz, as well as other fits proposed by different authors:
\begin{enumerate}
    \item A pure power-law
    $$\sSFR \propto \Mstar^\alpha$$
    with $\alpha=-0.24$ that represents a fit to the Main Sequence of Star Forming galaxies proposed by \citet{2015RenziniPeng} to describe the mode of the distribution, below which quenched galaxies would remain without barely any ongoing star-forming activity (i.e. the paradigm of the bimodal interpretation).
    \item A broken power-law 
    $$\sSFR \propto \Mstar^{\alpha(\Mstar)}$$ 
    with $\alpha=-0.12$ for $\Mstar<10^{10}~\Msun$ and $\alpha=-0.73$ for \mbox{$\Mstar>10^{10}~\Msun$,} similar to the MS description provided by \citet{2015Lee}, aimed to describe the turnover at a critical mass $\sim 10^{10}~\Msun$.
    \item A Schechter function
    $$\sSFR = \sSFR_0 10^{(\alpha+1)(\lMstar-\log(\text{M}_0))}\exp\left({-10^{\lMstar-\log(\text{M}_0)}}\right),$$
    with $\sSFR_0=5.96 \times10^{-11}$ yr$^{-1}$,$~\log(\text{M}_0)=11.03,~\alpha=-1.35$ \citep{2007Salim}, explicitly fit to the mode of the conditional probability distribution.
    \item A polynomial fit
    $$\log(\sSFR) = -10.39 -0.479(\log(\Mstar) -10) -0.098(\log(\Mstar)-10)^2,$$
    proposed by \citet{2017Eales} to describe the whole galaxy population.
\end{enumerate}
We find that all the proposed fits are able to correctly reproduce the mode of the probability distribution, and they also provide a good description of the average \sSFR\ at any given stellar mass.
However, previous attempts cannot, by construction, account for the extended power-law tails quantitatively traced by the percentiles of the distribution.

The shape of the distribution in the $\Mstar-\sSFR$ plane \newchange{is roughly consistent with an evolutionary scenario where} low-mass galaxies tend to be found in a more `primitive' state, consistent with higher gas fractions and lower metallicities \citep[see e.g.][]{2015Ascasibar}, whereas high-mass galaxies tend to be more `chemically evolved'.
One possible interpretation is that, below $\sim 10^{10}~\Msun$, most galaxies live in a quasi-steady state where SFR and gas infall are approximately balanced.
Only systems in dense environments, where infall is partially or entirely suppressed (or cold gas is actually removed from the galaxy) display low specific star formation rates.

For intermediate masses, $\lMstar\gtrsim 10$, the mode, mean, or any percentile of the distribution become increasingly lower.
This fact can be understood as some sort of gas-consumption phase where star-formation decays due to the depletion of the cold gas reservoir.
\newchange{At variance with smaller galaxies, the bath-tub model might no longer represent the average population, either because infall/cooling becomes statistically less efficient and/or the conversion of gas into stars becomes more efficient.
Alternatively, the equilibrium state might depend systematically on stellar mass.}
Determining the precise shape of the probability distribution in this mass regime is extremely important in order to constrain the characteristic time scales of the physical mechanisms involved \citep[see e.g.][]{2019Phillipps}, but we would like to argue that we do not see evidence of a sharp transition between two discrete states associated to a particular event.

First, we do not observe any strong bimodality in the SDSS data, \nnewchange{in contrast with the bimodal behaviour of GAMA. 
Once again, we support the scenario of a unimodal distribution given the accuracy of the current data.}
Although this does not preclude the occurrence of `quenching' events in some galaxies, especially in dense environments, we do not think that they are \emph{necessary} to explain the evolution of the overall population.
Furthermore, even if the conditional probability turned out to be bimodal in this particular mass range, this would not necessarily imply the existence of sudden variations in the SFR.

In the `ageing' interpretation, there would simply be a smooth correlation between mass and the characteristic time scales for gas accretion and conversion into stars.
The relative scarcity of galaxies in the `Green Valley' would be merely a consequence of the non-linearity of the mapping between sSFR and colour \citep[e.g.][]{2017Eales}.
To some extent, a similar argument could also be applied to the star formation rate; if galaxies are smoothly distributed along a certain evolutionary track, the number of objects displaying a given sSFR value would be proportional to the time they spend there.
For any smooth star formation history that increases, reaches a maximum, and then decreases, all the galaxies that have not reached the peak yet would display sSFR of the order of the current age of the universe, whereas those well beyond the peak will accumulate at values close to zero.
Although one may regard `increasing' and `decreasing' SFR as two different `states', there are no significant variations (nor physical events) associated to the transition between both regimes.

Constraining the distribution at the high-mass end is also of the utmost importance.
In particular, the observed trends in this regime seem to indicate that old systems might tend to reach an asymptotic state with a minimum level of star formation, which could be due, for instance, to the contribution of recycled gas from the old stellar population \citep[e.g.][]{2019SalvadorRusinol, 2020Eales}.
If such asymptotic scale were realised in nature, it would not be surprising to find a bimodality in the distribution of \sSFR, much similar to the one observed in the colour-magnitude diagram \citep[e.g][]{2017Eales}.
There would be plenty of galaxies accumulating near the asymptotic state (the reddest possible colour, corresponding to a $\sim10$~Gyr-old population, or the \sSFR\ level associated to gas recycling), but this would by no means imply a sharp transition between two discrete states.
With all the information that is currently available, both the `ageing' \citep[e.g.][]{2013Gladders,2015Abramson,2015Casado} and the `quenching' \citep[e.g.][]{2010Peng} scenarios are perfectly compatible with the data.

\begin{figure*}
    \centering
\includegraphics[width=\linewidth]{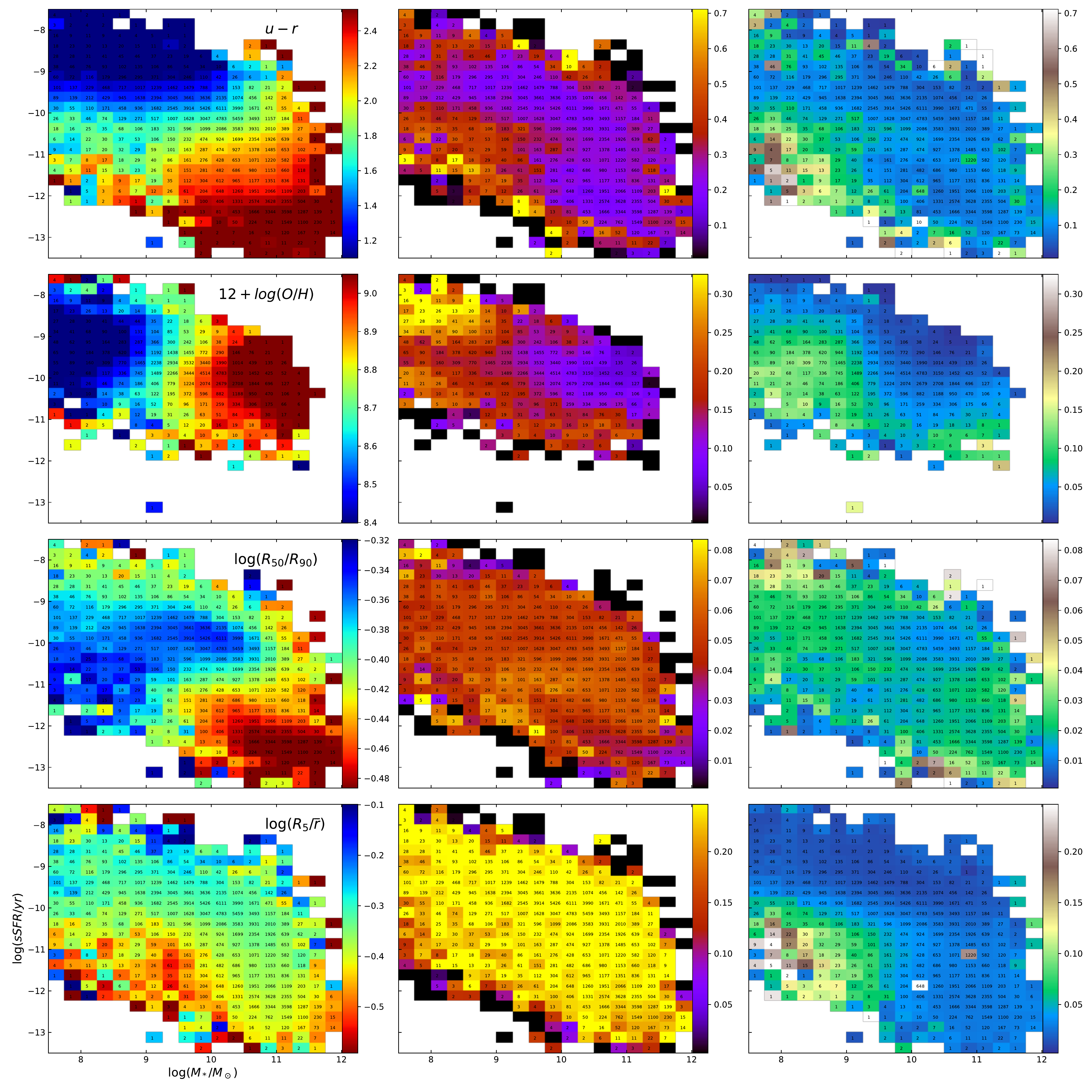}
\caption{$\sSFR-\Mstar$ plane of SDSS galaxies. 
Each row shows coloured bins based on colour $u-r$, oxygen abundance $12+\log(O/H)$, inverse concentration index $R_{50}/R_{90}$ and  the environment indicator $\eta$. Left, mid and right columns represent the mean value, dispersion and average observational uncertainties within each bin with numbers indicating the number of galaxies taken into account.}
\label{fig:extra_params}
\end{figure*}

\subsection{Degrees of freedom}
\label{subsec:dof}

In principle, galaxies could display any combination of stellar mass and star formation rate.
Although it has been proposed that a purely stochastic correlation may also arise as a consequence of the central limit theorem \citep{2014Kelson}, it is well known that halo mass, accretion history, and star formation are closely related \citep[e.g.][]{2013Behroozi}, and therefore it is not surprising that $M_*$ and sSFR are not independent quantities.
As a matter of fact, the existence of strong correlations between many of the physical properties of galaxies has led to the suggestion that they can be fully described in terms of just one \citep[e.g][]{2008Disney}, two \citep[e.g.][]{1995Connolly, 2006Smolcic, 2011AscasibarSanchezAlmeida}, or three \citep[e.g.][]{2004Yip} free parameters.

One of them could represent a physical scale (such as the stellar, baryon, or halo mass, the effective or half-light radius, the escape velocity, etc.), and another could be any intensive quantity related to the evolutionary state (e.g. sSFR, stellar-to-gas fraction, stellar or gas metallicity).
All these ratios are known to correlate with the scale parameter to some extent, and many such `scaling relations' have been reported in the literature.

If there was only one degree of freedom, all these correlations would be extremely tight, and galaxies would be arranged along an infinitely thin curve in the $\Mstar-\sSFR$ plane.
Otherwise, they would define a hypersurface (for two independent parameters) or a hypervolume (for three or more degrees of freedom) in the space of all possible physical properties.
It would be very unlikely that these surface or volume would project as a narrow curve in the $M_*-sSFR$ plane, and therefore one would expect, consistently with our results, that galaxies display an (uneven) distribution of masses and star formation rates.

As nicely summarised by \citet{2015Abramson}, there are three different interpretations of the width of the probability distribution $\deriv{p(SFR|M_*)}{sSFR}$.
One possibility \citep[e.g.][]{2010Peng, 2013Behroozi, 2015RenziniPeng} is that there is only one degree of freedom; all galaxies form their stars exactly according to a given function $\SFR(\Mstar, t)$, and thus the stellar mass at the present time suffices to reconstruct the full history of any galaxy across cosmic time.
Deviations from the `main sequence' are attributed to random fluctuations associated to physical processes acting on short time scales whose amplitude would explain the width of the observed distribution.

We think that there are strong arguments against that scenario.
The magnitude of random fluctuations of the star formation rate on scales of the order of $\sim 30$~Myr is severely constrained by the tightness of the distribution of the equivalent width of the H$\alpha$ emission line with respect to other tracers of star formation on longer timescales \citep[e.g.][]{2015Casado, 2019WangLilly}.
Models addressing the star formation histories of individual galaxies including stochastic processes also find that they are correlated on timescales of the order of $0.1-1$~Gyr \citep[e.g.][]{2015MunozPeeples, 2019Caplar} or even longer \citep[e.g.][]{2019Matthee}.
Moreover, galaxies have been shown to display vastly different (yet fairly smooth) star formation histories at fixed stellar mass \citep[e.g.][]{2015Abramson, 2016Abramson, 2016Dressler, 2017Oemler} that are strongly correlated with morphological type \citep[e.g.][]{2015GonzalezDelgado}.

According to \citet{2015Abramson}, one may distinguish random fluctuations correlated on intermediate scales from systematic differences in the star formation history across the entire life of the galaxy.
In the former case, there would be many different paths leading to the current state $(\Mstar,  \sSFR)$ of the galaxy, and therefore other physical properties may take arbitrary values, whereas for coherent variations there would only be one possible path, and any other property of the galaxy would be univocally determined.

We would like to argue that this picture is perfectly valid if galaxies are fully described by two degrees of freedom.
If there were other relevant factors (e.g. morphology, environment), and they were not perfectly correlated with the star formation history, correlations of other physical properties with the sSFR at fixed mass would wash out.
Thus, the lack of correlation would imply that other variables play a role, but it would not be possible to distinguish between random fluctuations and a systematic physical effect.
On the other hand, a strong correlation between any property of a galaxy with the sSFR at fixed mass would rule out that this property is strongly affected by random fluctuations, but it would not prove that such fluctuations do not exist.

Trying to shed light into this question, we have studied the degree of correlation of different observables with the location of a galaxy on the $\sSFR-\Mstar$ plane.
The $u-r$ colour is determined by a mixture of the fraction of young stars, the metal content of the galaxy, and dust extinction (involving dust mass and geometry); the oxygen abundance, 12+log(O/H), is another tracer of the evolutionary state of the galaxy; the inverse concentration index (ratio between the 50-percent and 90-percent total-light Petrosian radii, $R_{50}/R_{90}$) is proxy of galaxy morphology; and a dimensionless quantity
\begin{equation}
 		\eta = \frac{R_5}{\overline{r}}
\end{equation}	
defined as the three-dimensional distance $R_5$ to the 5th nearest-neighbour within a shell of 30 Mpc thickness ($\Delta z = 0.005$) normalized to the average value
\begin{equation}
\overline{r}(z) =
\left[
\frac{
	\left(\frac{d_L(z)}{(1+z)^2} + 15~\text{Mpc}\right)^3
	-\left(\frac{d_L(z)}{(1+z)^2} - 15~\text{Mpc}\right)^3
}{N_{gal}(z)}
\right]^{1/3}
\end{equation}
expected for a homogeneous distribution of $N_{gal}(z)$ galaxies at redshift $z$, with $d_L(z)$ denoting the comoving luminous distance, provides an indicator for environment.

The average value of these quantities as a function of $sSFR$ and $\Mstar$ is represented on the left column panels of Fig.~\ref{fig:extra_params}.
In addition, mid column panels represent the dispersion $\sqrt{\langle (\log x)^2 \rangle - \langle \log x \rangle^2}$ within each bin, that includes the contribution of intrinsic scatter in galaxy properties as well as measurement errors.
For that reason, the average observational uncertainty in every bin is shown on the right column panels. Comparing both quantities is useful to check whether the dispersion is real or instead it may be dominated by the uncertainties.
On the other hand, the range of the color maps in the left column corresponds to the 16th to the 84th percentile for every quantity, while for the mid and right panels it saturates at the characteristic dynamic range given by $p_{84}-p_{16}$.

The $u-r$ colour index displays a clear correlation with $sSFR$, where most of galaxies belonging to the BC tend to be located above a certain threshold, and they usually have lower stellar masses than red galaxies.
As one moves towards the massive, red end, a secondary correlation with mass appears, presumably as a consequence of the mass-metallicity relation (metal-rich populations tend to display redder colours), and the isocontours are no longer horizontal but diagonal lines.
Conversely, metallicity is mostly correlated with stellar mass (vertical isocontours), with only marginal evidence of a non-linear dependence on \sSFR.
The physical meaning of the `Blue Cloud' would be a somewhat arbitrary threshold in sSFR, whereas the `Red Sequence' would be more affected by metallicity.
As pointed by e.g. \citet{2017Eales}, the colours of the `Green Valley' may only be observed for a very narrow range of sSFR, which explains the low number of galaxies that can be found in that region of the colour-magnitude diagram.

In both cases, the variation over the $\Mstar-\sSFR$ plane is statistically significant, much larger than the dispersion within each bin.
The latter, in turn, is always above the typical level of the reported observational errors, meaning that the dispersion is real, at the level of $\sim 0.2-0.3$~dex in both cases, giving the impression that $sSFR$ and $\Mstar$ are only able to roughly sketch the chemical composition and colour of a given galaxy (at least, no more accurately than that level).

As it is well known, morphology also correlates with \Mstar\ and \sSFR.
However, the relation is much more complex than in the case of colours and oxygen abundance.
The dispersion is also a little bit higher, enough to blur any tight morphological sequence in this plane.
Likewise, environment follows a trend such that galaxies with higher \sSFR\ (at fixed stellar mass) tend to live in less dense regions.
Once again, we do not detect any evidence of a sharp transition indicating a possible environmental threshold in terms of $\eta$.
We find that the environment of a galaxy is even less constrained than morphology by its location on the $\Mstar-\sSFR$ plane, since the intrinsic dispersion is comparable to the entire dynamic range of the proxy $\eta$.
In both cases, the dispersion can clearly not be attributed to observational errors.

These results evidence the importance of considering (at least) a third parameter in order to fully describe a galaxy in addition to its scale (traced by e.g. stellar mass) and evolutionary state (e.g. specific star formation rate).
Given the dispersion present at every panel, environment seems to be the best candidate as an independent (most informative) parameter, as it is the worst constrained by $sSFR$ and $\Mstar$.

\section{Conclusions}
\label{sec:conclusions}

In this work we have used observational measurements based on the SDSS and GAMA surveys to characterise the distribution of local galaxies in the $\sSFR-\Mstar$ plane in terms of the conditional probability \pp\ of the specific star-formation rate at given stellar mass.
Our results highlight that the galaxy distribution is intrinsically bidimensional.
The so-called `Main Sequence of Star-Forming galaxies' roughly corresponds to the mode in \sSFR\ as a function of \Mstar.
However, a 'tight' sequence of the form $sSFR(\Mstar)$ \nnewchange{and a completely detached population of passive galaxies at much lower levels of star formation activity may not provide the best possible description of the galaxy demographics}.

The conditional probability is broad, with a dispersion of the order of $0.6-1$~dex when no attempt is made of excluding `passive' galaxies.
Furthermore, it is not well fitted by a \newchange{single} log-normal distribution because it features extended tails in both directions.

\newchange{In this work we have compared the performance of fitting a double log-normal distribution, understood as a representative model of the bimodal paradigm, against the results obtained from fitting a unimodal distribution with asymptotic power-law tails, intended to be representative of the whole (single) galaxy population.
We find that both approaches are statistically compatible with the data to the level of current uncertainties, with SDSS data being slightly better fitted by our unimodal prescription whereas GAMA results are better reproduced by the two log-normal distribution.
Our most important result is therefore that the single-population description should not be discarded with the current accuracy of the data.}

Regarding the physical interpretation of the observed distribution, the correlation between the location in the $\sSFR-\Mstar$ plane and other galaxy properties such as colour, metallicity, morphology or environment provides additional evidence that the value of the instantaneous SFR measured at the present time is not a random fluctuation around a `Main Sequence' driven entirely by stellar mass.

We support the interpretation in terms of a single population of galaxies at different `ageing' stages, in contrast to the bimodal paradigm of 'active' and 'passive' galaxies separated by `quenching' processes.
Although we find marginal evidence in GAMA that the conditional probability $p$ might be bimodal in the mass range \mbox{$\sim 10^{10} - 10^{11}~\Msun$,} it mostly seems to be continuous and unimodal for most stellar masses.
Even if some bimodality was confirmed by more accurate observations, a more elaborate description of the conditional probability would be necessary, but it would not rule out the `ageing' interpretation.
\newchange{Conversely}, the lack of a clear separation between `star-forming' and `passive' systems is not incompatible with the `quenching' scenario.
A more detailed analysis of the star formation history on different timescales is required in order to address this issue.

In general, deeper observations are necessary to constrain the low-sSFR tail at all mass ranges.
For both dwarf and giant galaxies, the scarce number of objects (stemming from sensitivity and volume limits, respectively) prevents an accurate exploration of the probability distribution.
In the intermediate regime, better statistics are also necessary to verify whether the dispersion in sSFR reaches a minimum near $\sim 10^9-10^{10}~\Msun$.
In addition to an improvement in the quantity and quality of data, a careful assessment of the statistical and systematic uncertainties associated to the reconstruction of the SFH must be carried out.

Finally, let us note that, albeit the characterisation of the distribution in the $\sSFR-\Mstar$ plane is an important step forward in our understanding of the physical properties of galaxies, our results also show that additional parameters are required to obtain a full picture of galaxy evolution.
In our opinion, morphology and environment would be the most natural candidates, with the latter being slightly favoured by the weaker correlation with mass and star formation rate.


\section*{Acknowledgements}

The authors are indebted to the anonymous referee for a rigorous and constructive report, which has led to a major reformulation of the original manuscript.
The authors also want to thank M. Gavil\'an for her support and inestimable help along this work.

This work has been financially supported by the Spanish Government project ESTALLIDOS: AYA2016-79724-C4-1-P (Ministerio de Ciencia, Innovaci\'on y Universidades, Spain).

Funding for the Sloan Digital Sky Survey IV has been provided by the Alfred P. Sloan Foundation, the U.S. Department of Energy Office of Science, and the Participating Institutions. SDSS-IV acknowledges
support and resources from the Center for High-Performance Computing at
the University of Utah. The SDSS web site is www.sdss.org.

SDSS-IV is managed by the Astrophysical Research Consortium for the 
Participating Institutions of the SDSS Collaboration including the 
Brazilian Participation Group, the Carnegie Institution for Science, 
Carnegie Mellon University, the Chilean Participation Group, the French Participation Group, Harvard-Smithsonian Center for Astrophysics, 
Instituto de Astrof\'isica de Canarias, The Johns Hopkins University, Kavli Institute for the Physics and Mathematics of the Universe (IPMU) / 
University of Tokyo, the Korean Participation Group, Lawrence Berkeley National Laboratory, 
Leibniz Institut f\"ur Astrophysik Potsdam (AIP),  
Max-Planck-Institut f\"ur Astronomie (MPIA Heidelberg), 
Max-Planck-Institut f\"ur Astrophysik (MPA Garching), 
Max-Planck-Institut f\"ur Extraterrestrische Physik (MPE), 
National Astronomical Observatories of China, New Mexico State University, 
New York University, University of Notre Dame, 
Observat\'ario Nacional / MCTI, The Ohio State University, 
Pennsylvania State University, Shanghai Astronomical Observatory, 
United Kingdom Participation Group,
Universidad Nacional Aut\'onoma de M\'exico, University of Arizona, 
University of Colorado Boulder, University of Oxford, University of Portsmouth, 
University of Utah, University of Virginia, University of Washington, University of Wisconsin, 
Vanderbilt University, and Yale University.

GAMA is a joint European-Australasian project based around a spectroscopic campaign using the Anglo-Australian Telescope. The GAMA input catalogue is based on data taken from the Sloan Digital Sky Survey and the UKIRT Infrared Deep Sky Survey. Complementary imaging of the GAMA regions is being obtained by a number of independent survey programmes including GALEX MIS, VST KiDS, VISTA VIKING, WISE, Herschel-ATLAS, GMRT and ASKAP providing UV to radio coverage. GAMA is funded by the STFC (UK), the ARC (Australia), the AAO, and the participating institutions. The GAMA website is http://www.gama-survey.org/ . 

This research has made use of NASA's Astrophysics Data System.

\section*{Data availability}

The data underlying this article are available at \url{https://www.sdss.org/dr16/} for the SDSS survey and \url{http://www.gama-survey.org/dr3/} for GAMA. Additional  data generated by the analyses in this work are available upon request to the corresponding author.  



\bibliographystyle{mnras}
\bibliography{bibliography} 


\bsp	
\label{lastpage}
\end{document}